\documentclass[a4paper,11pt]{article}
\pdfoutput=1
\usepackage{jheppub}
\usepackage{amsmath,graphicx,slashed,multirow,color,ulem}
\usepackage{graphicx,amssymb,bm,wasysym,multirow}
\usepackage[utf8]{inputenc}
\usepackage[numbers,sort&compress]{natbib}

\def \beq{\begin{equation}}
\def \eeq{\end{equation}}

\title{
Understanding forward $B$ hadron production
}

\author[a]{Rhorry Gauld}
\affiliation[a]{ETH Zurich, Institut fur theoretische Physik, Wolfgang-Paulistr. 27, 8093, Zurich, Switzerland}

\date{March 2017}

\abstract{
The LHCb collaboration has recently performed a measurement of the production rate of inclusive 
$B$ hadron production ($pp\to BX$) at both 7 and 13~TeV centre-of-mass (CoM) energies. 
As part of this measurement, the ratio of these two cross section measurements has been
presented differentially in $B$ hadron pseudorapidity within the range of $\eta_B \in [2.0,5.0]$.
A large tension ($4\sigma$) is observed for the ratio measurement in the lower pseudorapidity 
range of $\eta_B \in [2.0,3.0]$, where the data is observed to exceed theoretical predictions, 
while consistency is found at larger $\eta_B$ values. This behaviour is not expected within perturbative QCD, 
and can only be achieved by introducing ad-hoc features into the structure of the non-perturbative gluon PDF 
within the region of $x\in[10^{-3},10^{-4}]$. Specifically, the gluon PDF must grow extremely quickly 
with decreasing $x$ within this kinematic range, closely followed by a period of decelerated growth. 
However, such behaviour is highly disfavoured by global fits to proton structure. Further studies of the 
available LHCb $B$ and $D$ hadron cross section data, available for a range of CoM energies, 
indicate systematic tension in the (pseudo)rapidity region of $[2.0,2.5]$.
}

\emailAdd{rgauld@phys.ethz.ch}

\begin{document}
\maketitle
\flushbottom

\section{Introduction}
The LHCb collaboration has recently presented measurements of inclusive $B$ hadron production in 
$pp$ collisions at 13 and 7~TeV centre-of-mass (CoM) energies~\cite{Aaij:2016avz}, defined through 
the process $pp\to BX$.
The cross section measurements are reported differentially with respect to 
$B$ hadron pseudorapidity ($\eta_B$) within the range $\eta_B \in [2.0,5.0]$,
and inclusively with respect to transverse momentum ($P_T^{B}$).
In addition, the ratio of the differential cross section measurements at these
two CoM energies has also been presented. 

The motivation for considering the ratio of heavy quark cross section measurements
is that many sources of (otherwise overwhelming) theoretical and experimental 
uncertainty, which are highly correlated at different CoM values, partially cancel 
in the ratio. At the same time, the ratio is still sensitive to the shape of the gluon 
parton distribution function (PDF) at both small and large values of Bj\"orken-$x$ 
($x$)~\cite{Gauld:2015yia,Cacciari:2015fta}, since typically different values of $x$ are 
probed within a fixed kinematic region at different CoM values.
Consequently, it is possible to include the heavy quark data at the level of the
ratio into a global analyses of proton structure, improving the description of the gluon PDF. 
This method was recently applied~\cite{Gauld:2016kpd} to the double differential 
$D$ hadron ratio data provided by LHCb~\cite{Aaij:2016jht,Aaij:2013mga,Aaij:2015bpa}.

It therefore comes as a quite a surprise that significant tension is observed 
for the $B$ hadron ratio data with respect to the corresponding theoretical 
predictions. In particular, the data is observed to exceed ($\approx4\sigma$) 
the predictions in the range of $\eta_B \in [2.0,3.0]$, while agreement is found 
for the more forward region of $\eta_B \in [3.0,5.0]$. This behaviour
is unexpected for the following reasons.
\begin{itemize}
\item Firstly, while $B$ and $D$ hadron predictions typically probe different 
values of $x$ and $Q^2$ of the input PDFs, there are kinematic regions where 
the two predictions are highly correlated. No tension is observed for the most 
precise (13/5~TeV) $D$ hadron ratio measurement in these regions~\cite{Aaij:2016jht}.
\item Secondly, a striking feature of the $B$ hadron data is that the ratio is observed to
decrease with increasing $B$ hadron pseudorapidity, which would indicate the presence 
of a region of accelerated then decelerated growth of the gluon PDF at values of $x \in [10^{-3},10^{-4}]$ 
and $Q^2 \sim 50{\rm~GeV^2}$. This is not a feature of DGLAP evolution, so such a structure
would have to be present in the non-perturbative gluon PDF. However, measurements of the
heavy quark (charm and beauty) structure functions $F_2^{qq}(x,Q^2)$ at HERA~\cite{Abramowicz:2014zub} 
do not find such a feature in this $x$ range, where this sort of effect should be more pronounced since the relevant
data is at lower $Q^2$ values.
\end{itemize}

\noindent
The purpose of this work is to perform detailed studies of 
the available forward $B$ hadron production data to better 
understand the possible origin of the observed deviation.
The remainder of this paper is organised as follows.
In \S~\ref{Theory}, the theoretical set-up for providing $B$ hadron production predictions 
are discussed, and the kinematics relevant for $B$ hadron production within the LHCb acceptance 
are studied.
In \S~\ref{Sigma}, the available LHCb $B$ hadron cross section data is studied at the level
of both the absolute and normalised cross sections at both
7~\cite{Aaij:2013noa} and 13~TeV~\cite{Aaij:2016avz} CoM energies.
After studying the cross section data, the ratio of the 13 and 7~TeV cross section measurements 
is studied in \S~\ref{Ratio}. In addition to studying the differential ratio as measured by LHCb, a kinematically 
`shifted' ratio is introduced which provides direct sensitivity to the growth of the low-$x$ gluon PDF. 
In \S~\ref{Consistency}, both the theoretical and experimental consistency of the 
LHCb $B$ hadron ratio data is considered. Firstly, the theoretical consistency of the data
is considered by comparing the experimentally extracted values for the growth of the gluon PDF
with those obtained with a toy model for PDFs. Secondly, correlations between the predictions for
$B$ and $D$ hadron are also considered, and the consistency of the $D$ hadron ratio measurements
are also discussed.
Finally, some general discussion and conclusions are provided in \S~\ref{Conclusions}.

\section{Theoretical set-up for forward $B$ hadron production}\label{Theory}
At the LHC, inclusive $B$ hadron production is dominated by the gluon-fusion
heavy quark pair production subprocess, and the predictions of the distributions of $B$ 
hadrons can be obtained by convoluting the partonic cross section for heavy quark pair production 
with input PDFs and the relevant heavy quark fragmentation functions.
The basis for the current state-of-the-art for differential cross section predictions is the next-to-leading order (NLO)
partonic cross section~\cite{Nason:1987xz,Nason:1989zy,Mangano:1991jk,Beenakker:1990maa,Beenakker:1988bq}, 
where predictions can be further improved by matching this massive
calculation to a parton shower or a massless calculation.
In the following, the theoretical set-up for providing $B$ hadron predictions will be provided.
In addition, the partonic kinematics relevant for forward $B$ hadron measurements in the LHCb
acceptance are also discussed. While the discussion is focussed towards $B$ hadron production,
the predictions for $D$ hadron production proceed in essentially the same way.

\subsection{General considerations}
In the current studies, predictions are provided at NLO accuracy matched
to a parton shower ({\sc\small NLO$+$PS}), which is achieved with the {\sc\small POWHEG} 
method~\cite{Nason:2004rx,Frixione:2007vw,Alioli:2010xd} to match the heavy
quark pair fixed-order calculation~\cite{Frixione:2007nw} with 
{\sc\small Pythia8}~\cite{Sjostrand:2007gs,Sjostrand:2014zea}.
As a baseline, the default {\sc\small Monash 2013} tune~\cite{Skands:2014pea} is used throughout.
For further details on the various approaches to $B$ (and $D$) hadron production, 
the reader is directed to~\cite{Cacciari:2012ny,Gauld:2015yia}, 
where a comparison of predictions obtained at {\sc\small NLO$+$PS} 
accuracy (including both {\sc\small POWHEG} and (a){\sc\small MC@NLO}~\cite{Frixione:2002ik,Alwall:2014hca} 
methods) and those obtained with the semi-analytic {\sc\small FONLL} 
approach~\cite{Cacciari:1993mq,Cacciari:1998it,Cacciari:2001td,Cacciari:2003zu,Cacciari:2005uk}
are performed. In addition, information on predictions obtained in the so-called {\sc\small GM-VFNS} 
scheme can be found in~\cite{Kniehl:2004fy,Kniehl:2005de,Kniehl:2005mk,Kniehl:2005ej,Kneesch:2007ey,Kniehl:2009ar,Kniehl:2012ti}.
It is worth mentioning that while the calculation of next-to-NLO (NNLO) QCD corrections
for massive~\cite{Czakon:2012pz,Baernreuther:2012ws,Czakon:2013goa} (and massless~\cite{Currie:2016bfm}) 
quark pairs are complete, and results for top quarks distributions have been presented 
in~\cite{Czakon:2014xsa,Czakon:2015pga,Czakon:2015owf,Czakon:2016ckf}, 
the application of these results to $B$ (and $D$) hadron final states is not yet available.

\vspace{0.1cm}  \noindent\textbf{PDFs and $\alpha_s$}. 
For the input PDFs, the $n_f = 5$ fixed flavour number scheme (FFNS) PDF set 
NNPDF3.0 NLO $\alpha_s(m_Z) = 0.118$~\cite{Ball:2014uwa} with 1000 replicas is used, 
and accessed through the {\sc\small LHAPDF6} interface~\cite{Buckley:2014ana}.
The internal {\sc\small POWHEG} routines are altered to extract $\alpha_s$ 
from the grid provided with the PDFs, as oppose to using the internal $\alpha_s$ routines.	
As discussed in~\cite{Cacciari:1998it}, in such a set-up it is necessary to add compensation terms
to the evaluation of the differential cross section which account for the mismatch in the 
running of both $\alpha_s$ and PDF evolution with the fixed-order calculation --- which is performed
in a FFNS with $n_f = 3 (4)$ for charm (bottom) quark pair production. These compensation terms
are implemented in the {\sc\small POWHEG-HVQ} library.
The benefit of this approach is that the same PDFs are then used for both $B$ and $D$ hadron
predictions, and the contributions from the resummed charm quark PDF are included in the $B$ hadron
predictions.

\vspace{0.1cm} \noindent\textbf{Scale variation}.
The dynamical reference scale $(\mu_0)$ is set to the transverse mass
of the heavy quark in the underlying Born configuration ($m_T$). Scale variation is then 
performed by independently varying factorisation and renormalisation scales by a 
factor of two around the reference scale $\mu_0$ 
with the constraint $1/2 < \mu_R/\mu_F < 2$ (a 7-point scale variation).

\vspace{0.1cm}  \noindent\textbf{Input masses}. 
For the input heavy quark pole masses, the following choices for the central 
value and corresponding uncertainty are made
\beq
m_c = \left(1.50 \pm 0.20\right){\rm~GeV} \,,\qquad m_b = \left(4.75 \pm 0.25\right){\rm~GeV} \,.
\eeq
These values are consistent within uncertainties with the 
recommendations of the HXSWG~\cite{deFlorian:2016spz}.

\vspace{0.1cm}  \noindent\textbf{Fragmentation}.
In {\sc\small Pythia8}, the heavy quark fragmentation is performed
with the Lund-Bowler~\cite{Bowler:1981sb} approach --- see for example~\cite{Norrbin:2000zc}. 
The value of the fragmentation fractions, for example $f(b\to B^-)$, 
and distribution of the hadrons depends on the specifics of the particular tune. 
To investigate the dependence on the tune, distributions are
also computed with the {\sc\small 4C} tune~\cite{Corke:2010yf}. In addition,
the impact of manually varying the Lund-Bowler $B$ quark fragmentation variable of $r_b = 0.855$
in the default tune is also considered. It should be noted for normalised distributions and 
cross section ratios, the effects of varying fragmentation settings are negligibly small as compared
to scale and PDF uncertainties.

In the most recent LHCb measurement of inclusive $B$ hadron 
production~\cite{Aaij:2016avz}, the measurement is performed for the
sum of the (averaged over charge conjugate modes) following exclusive $B$ hadron modes: 
$\left\{B^0,B^+,B_s^0,\Lambda_b^0\right\}$.
In the case of $\Lambda_b^0$ production, a correction factor of 
$\delta = 0.25\pm0.10$ was also applied to account for undetected
$\Omega_b^-$ and $\Xi_b$ baryons.
To match this definition, the $B$ hadron final state is
also taken as the sum of these four exclusive final states (including a weight of 
1.25 for $\Lambda_b^0$ baryons) and the total sum of these contributions
is weighted such that $f(b\to B) = 1$. In essence,
\beq \label{eq:Bcontr}
\sigma(pp\to BX) = \frac{1}{2} \left( \sigma(B^0) + \sigma(B^+) + \sigma(B_s) + (1+ \delta) \sigma(\Lambda_b) + c.c.\right) \,.
\eeq
Unless distributions are shown for specific $B$ hadrons, this weighted
sum is always applied to the $B$ hadron final state.

\vspace{0.1cm}  \noindent\textbf{Total uncertainty}.
To evaluate the total uncertainty of the `{\sc\small NLO+PS}' predictions (labelled this way
in plots), the individual contributions from scale, $m_b$ and PDF variations are added 
in quadrature for both up and down variations as
\beq \label{eq:quad}
\delta {\rm Total} = \sqrt{\delta {\rm Scale}^2 + \delta {\rm PDF}^2 + \delta m_b^2} \,.
\eeq
In addition, a more conservative `Total uncertainty (linear)' will also be occasionally shown.
This is computed by adding the scale uncertainty linearly
with PDF and $m_b$ variations added in quadrature according to
\beq \label{eq:lin}
\delta {\rm Total~(linear)} = \delta {\rm Scale}+\sqrt{\delta {\rm PDF}^2 + \delta m_b^2} \,.
\eeq

\subsection{Kinematics}
The forward kinematic acceptance of the LHCb detector of $\eta \in [2.0,5.0]$
provides a unique opportunity to study heavy quark production in a kinematic regime
beyond the reach of the central LHC detectors. As the heavy quark pair production
process is dominated by gluon-fusion, such studies have the potential to probe the 
gluon PDF at both extremely small- and large-$x$ 
values~\cite{Gauld:2013aja,Zenaiev:2015rfa,Gauld:2015yia,Cacciari:2015fta,Gauld:2016kpd}.
The sensitivity of such measurements is easily understood by considering the PDF sampling
of the LO cross section
\begin{align} \label{eq:xsampling}
x_{1,(2)} \propto \frac{m_T}{\sqrt{S}}  \left( e^{(-)y_b} + e^{(-)y_{\bar{b}}} \right) \,,
\end{align}
where $\sqrt{S}$ is the hadronic CoM, and $y_{b,\bar{b}}$ are the outgoing heavy quark rapidity.
For both $B$ and $D$ hadron production, the LHCb detector has the capability to 
reconstruct hadrons from $p_T > 0$ (at small-$m_T$) and a large rapidities ($y_b \sim 4.5$) 
which provides sensitivity to low-$x$. Future measurements of $B/D$ hadron production
at large $p_T$ and $y_b$ also have the potential to probe the large-$x$ gluon PDF~\cite{Cacciari:2015fta}.

\begin{figure}[t!]
\begin{center}
\makebox{\includegraphics[width=0.49\columnwidth]{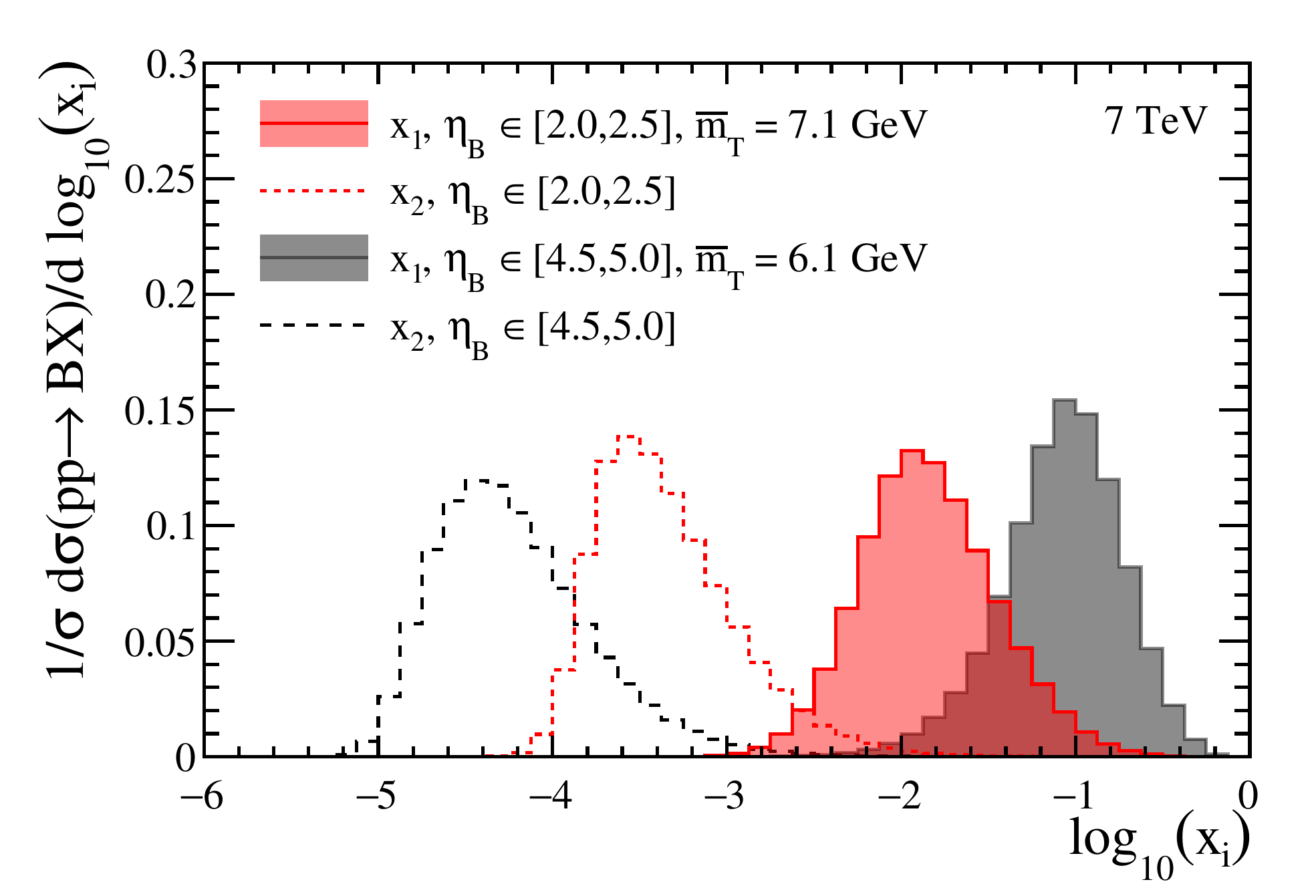}}
\makebox{\includegraphics[width=0.49\columnwidth]{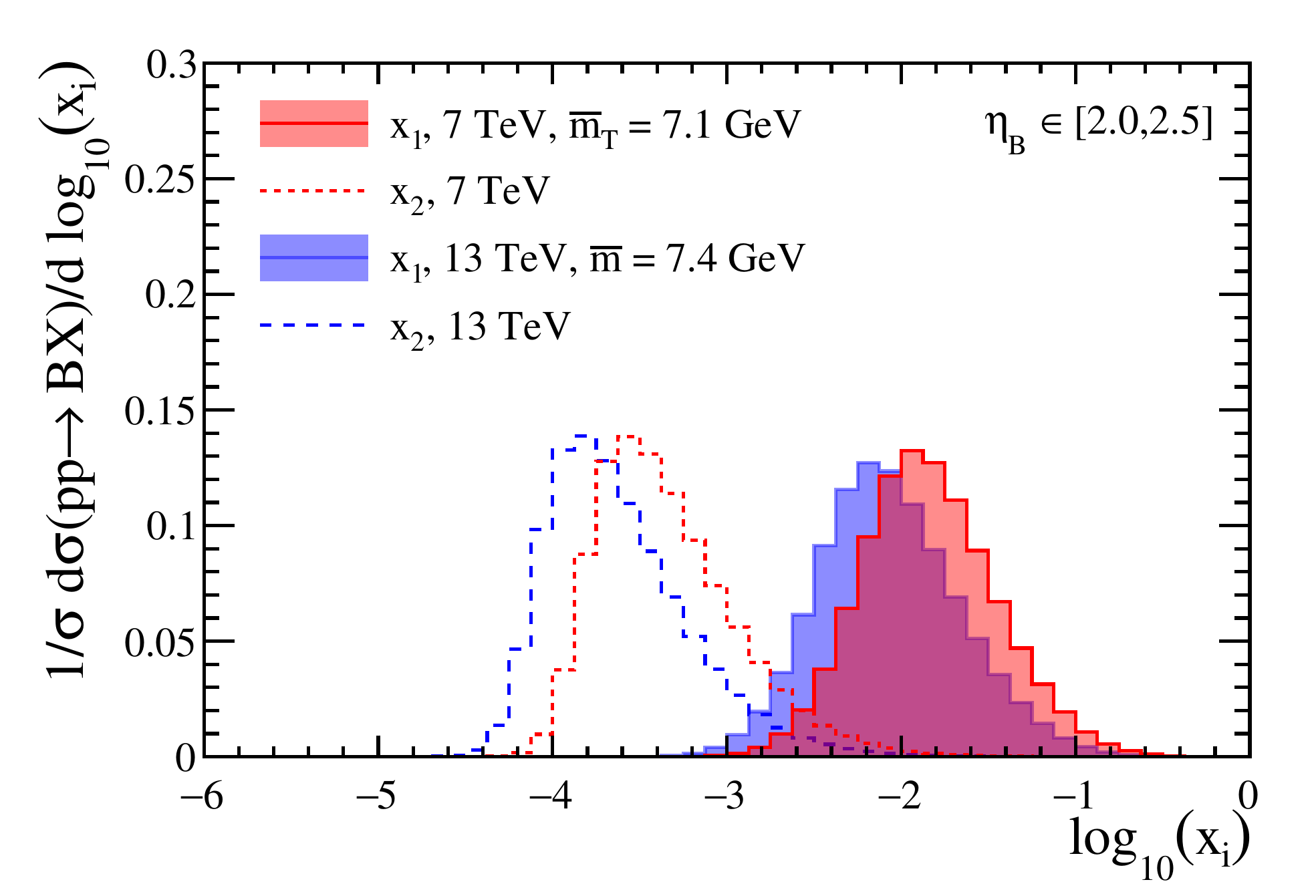}}
\end{center}
\vspace{0mm}
\caption{The LO $B$ hadron cross section as a function of the $x_{1,2}$ 
within specific pseudorapidity bins. Left: for varying choice of $\eta_B$ at 7~TeV. 
Right: for both 7 and 13~TeV with fixed $\eta_B \in [2.0,2.5]$.
}
\label{fig:x1x2}
\end{figure}

To understand the kinematic region relevant for forward $B$ hadron production,
the LO $B$ hadron cross section is shown in Fig.~\ref{fig:x1x2}, differentially in $x_{1,2}$.
In the left plot (7~TeV), the $B$ hadrons are required to be within either the pseudorapidity range 
$\eta_B \in [2.0,2.5]$ (red) or $\eta_B \in [4.5,5.0]$ (gray). Both of these kinematic
regions are accessed in the recent LHCb measurement. 
As expected from Eq.~(\ref{eq:xsampling}), increasing the value of the pseudorapidity requirement simultaneously increases (decreases)
the mean value of the $x_1 (x_2)$ PDF sampling region. With the requirement of $\eta_B \in [2.0,2.5]$, 
the mean PDF sampling occurs for $\bar{x}_1 \sim 1.4\cdot10^{-2}$ and $\bar{x}_2\sim 5\cdot 10^{-4}$ at
a scale of $Q^2\sim \overline{m}_T^2 \sim 50{\rm~GeV}^2$. In the right plot, the $B$ hadron cross section is shown both 
at 7 and 13~TeV, where the $B$ hadrons are required to be within the range of $\eta_B \in [2.0,2.5]$.
The mean value of the transverse quark mass ($\overline{m}_T$) which is probed for these selections 
is also highlighted. At fixed pseudorapidity, the mean values of the PDF sampling are decreased 
a factor of $\bar{x}_{i}^{13} \approx (7/13) \, \bar{x}_{i}^{7}$ when increasing $\sqrt{S}$ from 7$\to$13~TeV.
It is worth noting that the region of the gluon PDF which is probed for these kinematic selections
is well constrained (to a few~\%) by HERA DIS data~\cite{Aaron:2009aa,Abramowicz:1900rp,Abramowicz:2014zub}.

\section{(Normalised) $B$ hadron cross section data} \label{Sigma}
The purpose of this section is to perform a detailed study
of the shapes of forward $B$ hadron data available at 
both 7 and 13~TeV CoM energies~\cite{Aaij:2013noa,Aaij:2016avz}.
There are two distinct LHCb data sets which will be considered in the following 
analysis.
\begin{itemize}
	\item The first corresponds to the cross section measurement performed at both 7 and 13~TeV~\cite{Aaij:2016avz} 
	for $B$ hadrons reconstructed through the semi-leptonic decay modes $B\to D X \mu\nu$. The cross section
	ratio measurements, which will be discussed in the following Section, is performed with this data set.
	These measurements are presented differentially in $B$ hadron pseudorapidity, and 
	inclusively with respect to transverse momentum. The motivation for considering the
	semi-leptonic decays is that the relevant branching fractions are well known, which results in a 
	more precise determination of the absolute cross section rate. In contrast, the largest
	individual source of uncertainty for $B$ hadrons reconstructed through $B\to J/\psi X$ is
	associated to the branching fraction uncertainty.
	
	\item The second data set corresponds to the 7~TeV measurement of $B$ hadrons
	reconstructed exactly through the decay $B\to J/\psi X$~\cite{Aaij:2013noa}. This 
	measurement is performed for $B^+,B^0,B_s $  hadrons (and charge conjugate modes) where 
	all decay products are reconstructed, and both the transverse momentum 
	and rapidity dependence of $B$ mesons are accessed.
\end{itemize}

Before starting the comparison to data, it is worth mentioning that the experimental 
precision of these absolute cross section measurements is $\approx 10-20\%$. 
In contrast, the NLO accurate predictions for the absolute cross section have large uncertainties of 
$\approx 50\%$  --- for the most part dominated by scale uncertainties.
Consequently, a comparison of data to theoretical predictions at the level of the absolute 
cross section (although still important) is not particularly meaningful, since the overall normalisation
of the cross section is uncertain. Instead, as discussed in detail 
in~\cite{Zenaiev:2015rfa,Gauld:2015yia,Cacciari:2015fta,Gauld:2016kpd}, it is often
preferable to consider observables which are less sensitive to these scale uncertainties.
The general approach of this Section will be to perform the comparison to data 
both at the level of the absolute and normalised cross section.

\subsection{$B\to D\mu\nu$ cross section data (7 and 13~TeV)}
To begin, the recent forward $B$ hadron cross section measurement~\cite{Aaij:2016avz} 
is studied, where the $B$ hadrons have been identified
through the exclusive semi-leptonic decays $B\to D\mu\nu$.
As mentioned in the Introduction, this measurement is performed 
differentially in $\eta_B$ and inclusively in $p_T^B$, 
and the contributions from the sum of $B^+, B^0, B_s$ and $\Lambda_b$ hadrons 
(averaged over charge conjugate modes) as defined in Eq.~(\ref{eq:Bcontr}) 
are included\footnote{It may be possible to extend this measurement to 
reconstruct the $p_T^B$ dependence~\cite{Ciezarek:2016lqu}.}.

The strategy for performing a comparison to this data will be to first normalise the 
differential cross section data with respect to the integrated fiducial cross section 
measurement, defined as
\beq \label{eq:Beta}
\frac{1}{\sigma^{\rm fid.}_{\eta_B}} \frac{d \sigma(pp\to BX)}{d \eta_B} \,, \quad 
{\rm where~}\quad \sigma^{\rm fid.}_{\eta_B} = \int_{2.0}^{5.0} \frac{d \sigma(pp\to BX)}{d \eta_B} d \eta_B \,.
\eeq
The fiducial cross section data and corresponding theoretical predictions are summarised in Table~\ref{tab:sigfid},
where consistency (within large theoretical uncertainties) with the predictions is found for both 7 and 13~TeV measurements.
No correlation matrix has been provided for this $B$ hadron measurement,
and it is therefore assumed that the $\eta_B$-independent systematic uncertainties (as reported in Table~4 of~\cite{Aaij:2016avz})
are fully correlated between the fiducial and differential data points. For the study of a normalised 
cross section, it would be beneficial to have access to the experimental bin-by-bin correlations for the
cross section measurement.
\begin{table}[ht] 
\begin{center}
\begin{tabular}{r | c | c }
								& LHCb data [$\mu b$] 	& Theory [$\mu b$] 	\\ \hline
$\sigma_{\eta_B}^{\rm fid.}$~(7~TeV) 	& $72.0 \pm 0.3\,({\rm stat.})\pm 6.8\,({\rm sys.})$ & $56.7 ^{+28.6}_{-20.4}$	\\
$\sigma_{\eta_B}^{\rm fid.}$~(13~TeV) 	& $154.3 \pm 1.5\,({\rm stat.})\pm 14.3\,({\rm sys.})$ & $101.2 ^{+51.4}_{-39.8}$	\\
\end{tabular} \caption{Summary of 7 and 13~TeV measurements and predictions for the fiducial $B$ hadron cross
section $\sigma_{\eta_B}^{\rm fid.}$ within the LHCb acceptance.}
\end{center}
\label{tab:sigfid}
\end{table}

\noindent The motivation for normalising the cross section in this way is that
the large scale uncertainties in the absolute cross section are a result of varying 
the logarithmic scale dependence of the heavy quark transverse mass 
in the partonic cross section. However, this source of uncertainty primarily affects the overall 
normalisation of the cross section, and is highly-correlated between the neighbouring (pseudo)rapidity bins 
of the produced heavy quark. This observable is therefore theoretically more precise, and provides an 
important test of the shape of available data (rather than being overwhelmed by a normalisation
uncertainty). 

\begin{figure}[t!]
\begin{center}
\makebox{\includegraphics[width=0.49\columnwidth]{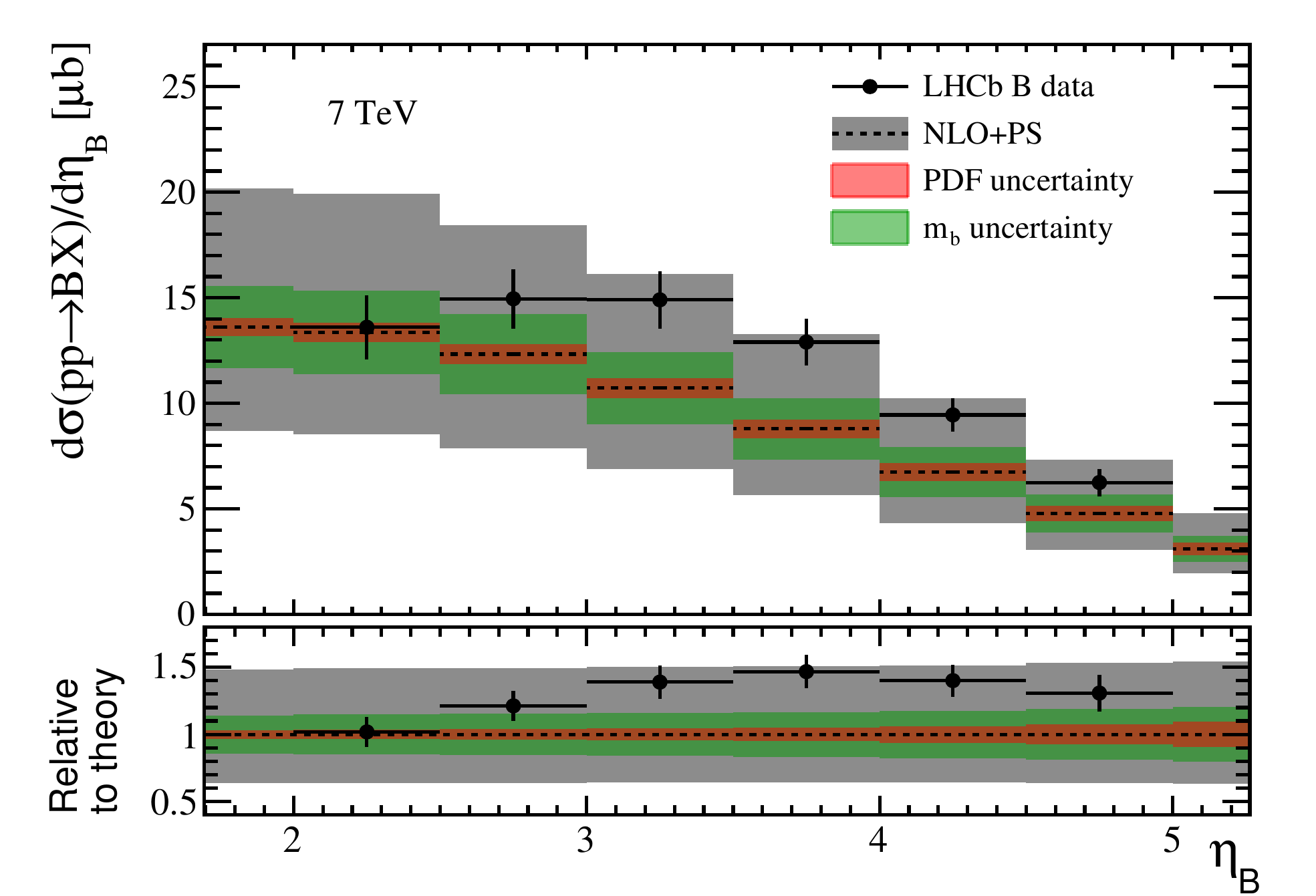}}  
\makebox{\includegraphics[width=0.49\columnwidth]{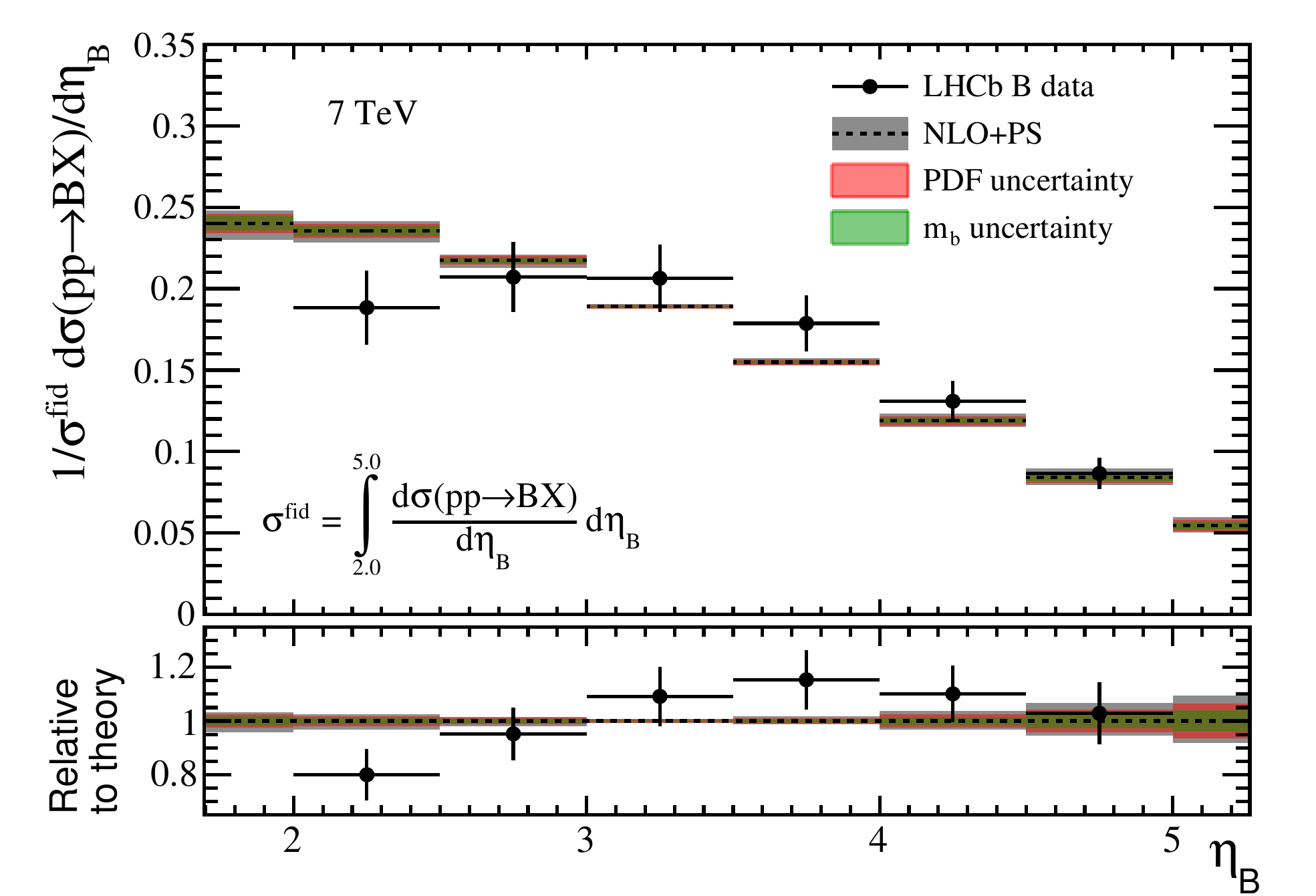}} \\
\end{center}
\vspace{0mm}
\caption{The absolute (left) and normalised (right) LHCb $B$ hadron cross section data at $\sqrt{S} = 7$~TeV. 
The theoretical uncertainty on the {\sc\small NLO$+$PS} accurate prediction is obtained as the 
sum in quadrature of the scale, PDF, and $m_b$ uncertainties.}
\label{fig:dsigdeta7}
\end{figure}

\begin{figure}[t!]
\begin{center}
\makebox{\includegraphics[width=0.49\columnwidth]{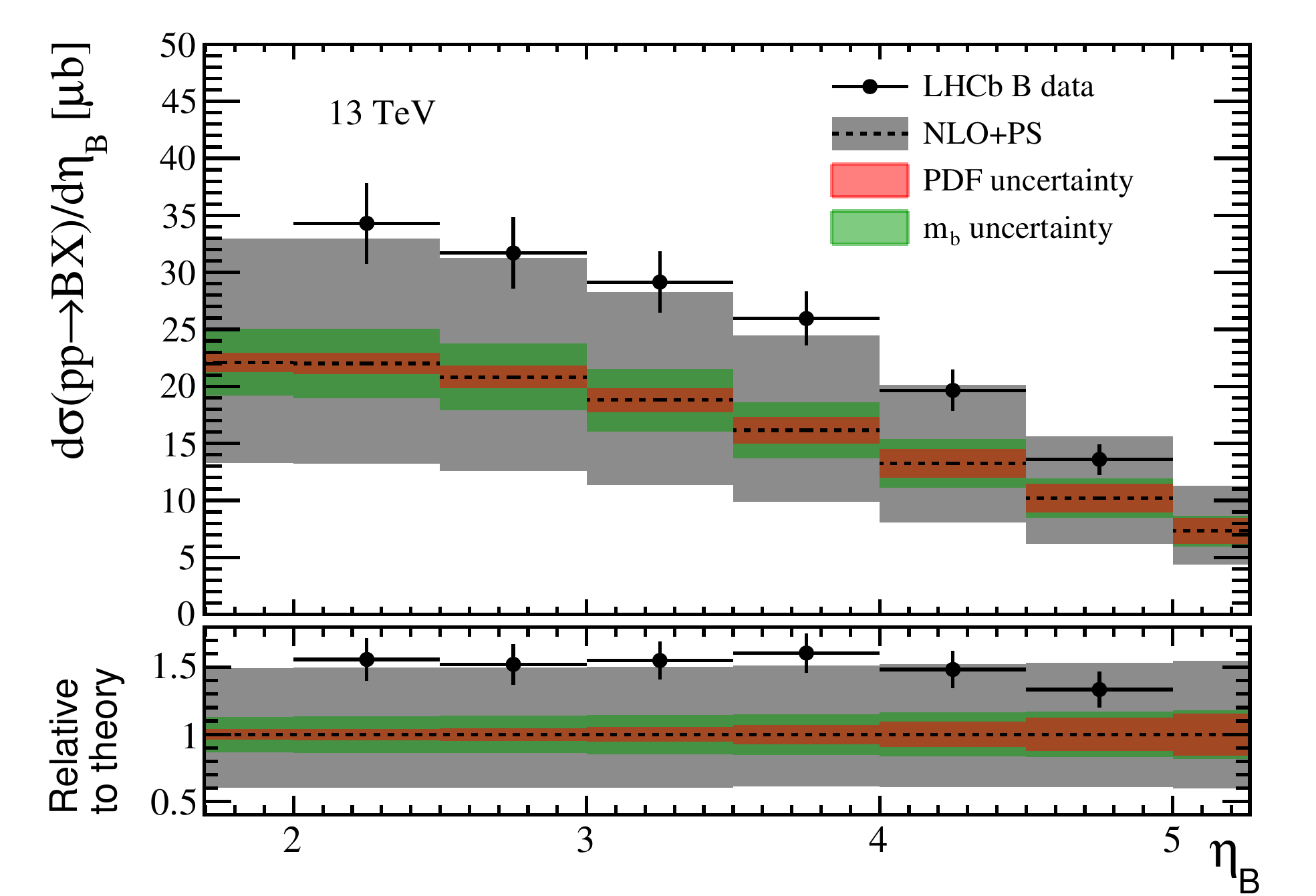}}  
\makebox{\includegraphics[width=0.49\columnwidth]{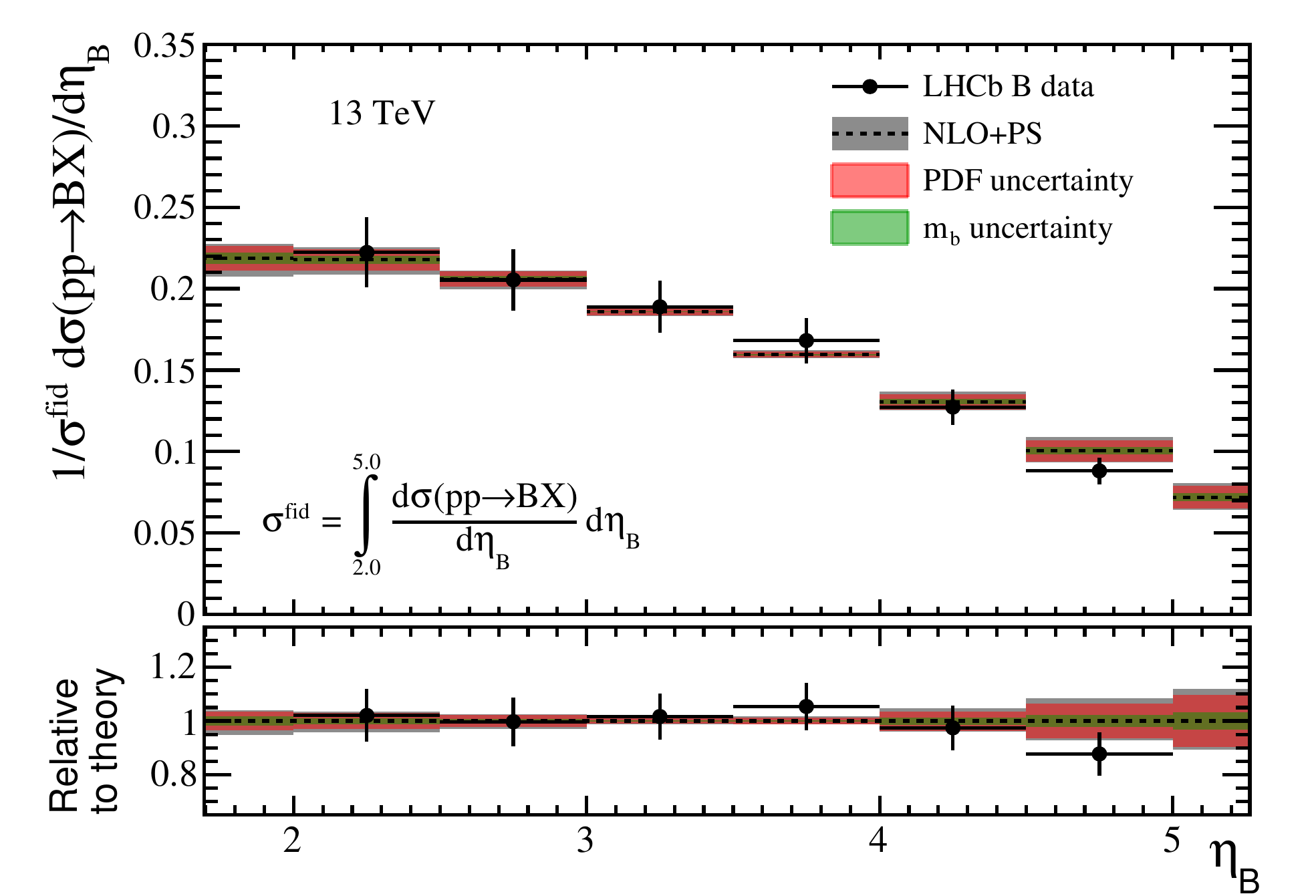}}
\end{center}
\vspace{0mm}
\caption{Same as Fig.~\ref{fig:dsigdeta7}, at $\sqrt{S}$ = 13~TeV.}
\label{fig:dsigdeta13}
\end{figure}

In Fig.~\ref{fig:dsigdeta7} and~\ref{fig:dsigdeta13}, 
the distributions for the absolute (left) and normalised (right) 
LHCb $B$ hadron cross section data is shown for 7 and 13~TeV respectively.
For each plot, the predictions and data are also shown normalised to the central theory
prediction in the lower panel. The total theoretical uncertainty
for the normalised cross section data is below 10\% while the absolute cross
section uncertainty is close to 50\%, demonstrating the above point. 
This approach also highlights an important feature of the data. For the case 
of the absolute cross section, the 7~TeV data tend to lie within the
(large) theoretical uncertainties while the 13~TeV data tend to lie at the upper
end of the theoretical scale uncertainties. At first glance, as the LHCb experiment reported,
this may indicate that ``The agreement with theoretical expectation is good
at 7~TeV, but differs somewhat at 13~TeV''.
However, as shown by the normalised distributions, this behaviour is not indicated. Actually, 
perfectly good agreement is found for the shape of the normalised 13~TeV cross section data, 
while the shape of the 7~TeV data is not as well described. This statement can be quantified 
by computing the $\chi^2/{\rm N_{dat}}$ for the data points with respect to the
central theory prediction, an approach which is justified for the normalised distribution as it has small 
theoretical uncertainties. This comparison gives
\beq \label{eq:chi2}
\chi_{\rm norm}^2/{\rm N_{dat}} {\rm (7~TeV)} = 8.2/6 \,,\qquad \chi_{\rm norm}^2/{\rm N_{dat}} {\rm (13~TeV)} = 2.9/6 \,.
\eeq
While the $\chi_{\rm norm}^2$ at 7~TeV is not particularly `bad', it is substantially worse
than that obtained at 13~TeV. The largest deviation is observed in the first bin, where the data is $2.1\sigma$ 
below the central theoretical prediction. It is worth mentioning that such a low value for the $\chi_{\rm norm}^2$ 
at 13~TeV may indicate that the experimental uncertainties are overestimated. This suggests that
it may be important to include bin-by-bin correlations when both normalising 
the data and computing the $\chi_{\rm norm}^2$.

Finally, it is worth commenting on the behaviour of the absolute cross section
at 13~TeV. In this case, it is observed that the absolute cross section tends to be 
on the upper end (although consistent within uncertainties) of the total NLO uncertainty band,
which is dominated by the scale uncertainty. A similar trend has also been
observed for $D$ hadron production within the LHCb acceptance at 5, 7 
and 13~TeV~\cite{Aaij:2016jht,Aaij:2013mga,Aaij:2015bpa}. 
This behaviour is entirely consistent with the observation that the NNLO corrections
to the absolute cross section for $t\bar{t}$ production (which, like $c\bar{c}$ and
$b\bar{b}$ pair production is also dominated by the gluon-fusion partonic subprocess) 
at the LHC are large and positive~\cite{Czakon:2015owf}.

\subsection{$B\to J/\psi X$ cross section data (7~TeV)}
In addition to the $\eta_B$ dependent cross section data, a double differential 
(in $p_T^B$ and $y^B$) cross section measurement was also performed
at 7~TeV~\cite{Aaij:2013noa}, where the $B$ hadrons have been reconstructed through the decay $B\to J/\psi X$.
It is useful to also consider the consistency of this data, to see if a similar trend is
observed for the normalised cross section data. In this case, comparisons are 
performed for both double and single ($p_T$-integrated) differential
cross section data.  When considering the rapidity distributions,
the following normalisation is applied
\beq \label{eq:singlediff}
\frac{1}{\sigma^{\rm fid.}_{y}} \frac{d \sigma(pp\to BX)}{d y^B} \,, \quad 
{\rm where~}\quad \sigma^{\rm fid.}_{y} = \int_{2.0}^{4.5} \frac{d \sigma(pp\to BX)}{d y^B} d y^B \,.
\eeq
Like Eq.~(\ref{eq:Beta}), this normalised distribution has the benefit that the 
uncertainty due to scale variation is highly correlated between numerator and 
denominator, since both are sensitive to similar values of $\overline{m}_T$.
To construct the experimental distributions, it is assumed that the branching 
ratio and luminosity uncertainty are fully correlated between bins. With this exception, 
the experimental uncertainties are added in quadrature as the bin-by-bin correlations are also 
not available for this measurement.
Both the experimental and theoretical rates for the fiducial cross section 
$\sigma^{\rm fid.}_{y}$ are reported in Table~\ref{tab:sigdy}.

\begin{table}[ht] \label{tab:sigdy}
\begin{center}
\begin{tabular}{r | c | c }
						& LHCb data [$\mu b$] 	& Theory [$\mu b$]	\\ \hline
$\sigma_y^{\rm fid.}(B^+)$	&	$38.9 \pm 0.3\,({\rm stat.})\pm 2.5\,({\rm sys.})\pm 1.3\,({\rm norm.})$ &	\multirow{2}{*}{$\frac{0.337}{f_B} (29.8 ^{+14.8}_{-10.3})$}	\\
$\sigma_y^{\rm fid.}(B^0)$		&	$38.1 \pm 0.6\,({\rm stat.})\pm 3.7\,({\rm sys.})\pm 4.7\,({\rm norm.})$ &			\\
$\sigma_y^{\rm fid.}(B_s)$	&	$10.5 \pm 0.3\,({\rm stat.})\pm 0.8\,({\rm sys.})\pm 1.0\,({\rm norm.})$ &	$\frac{0.092}{f_B} (8.12 ^{+4.05}_{-2.81})$		\\
\end{tabular} \caption{Summary of 7~TeV fiducial measurements and predictions for the fiducial $B$ hadron cross
section $\sigma_y^{\rm fid.}$ within the LHCb acceptance. The experimental uncertainties are statistical, systematic (including luminosity) and
normalisation due to branching fraction uncertainties. The over normalisation of the theoretical prediction depends on the value of the fragmentation fraction $f_B$ used for each $B$ hadron final state.}
\end{center}
\end{table}

\begin{figure}[t!]
\begin{center}
\makebox{\includegraphics[width=0.49\columnwidth]{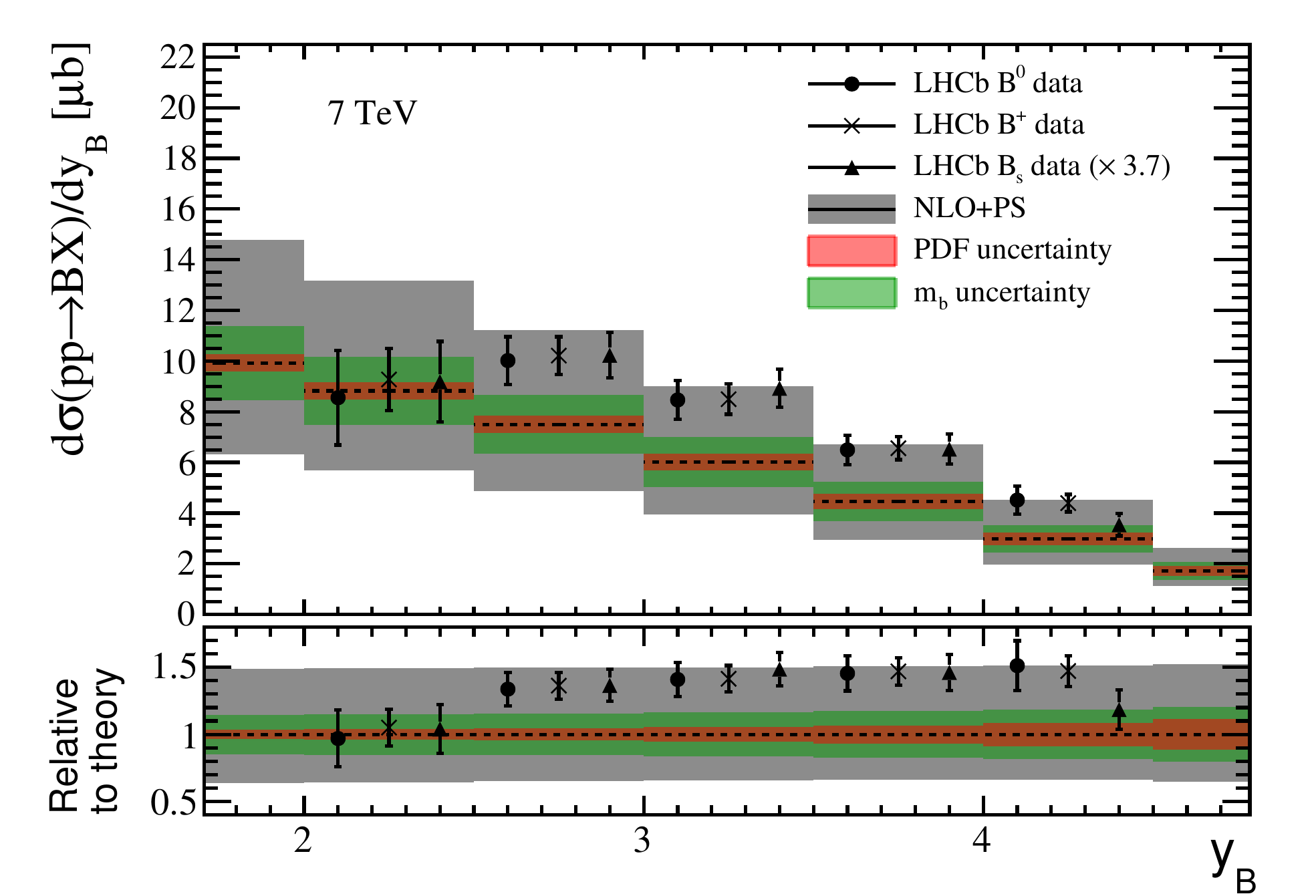}}
\makebox{\includegraphics[width=0.49\columnwidth]{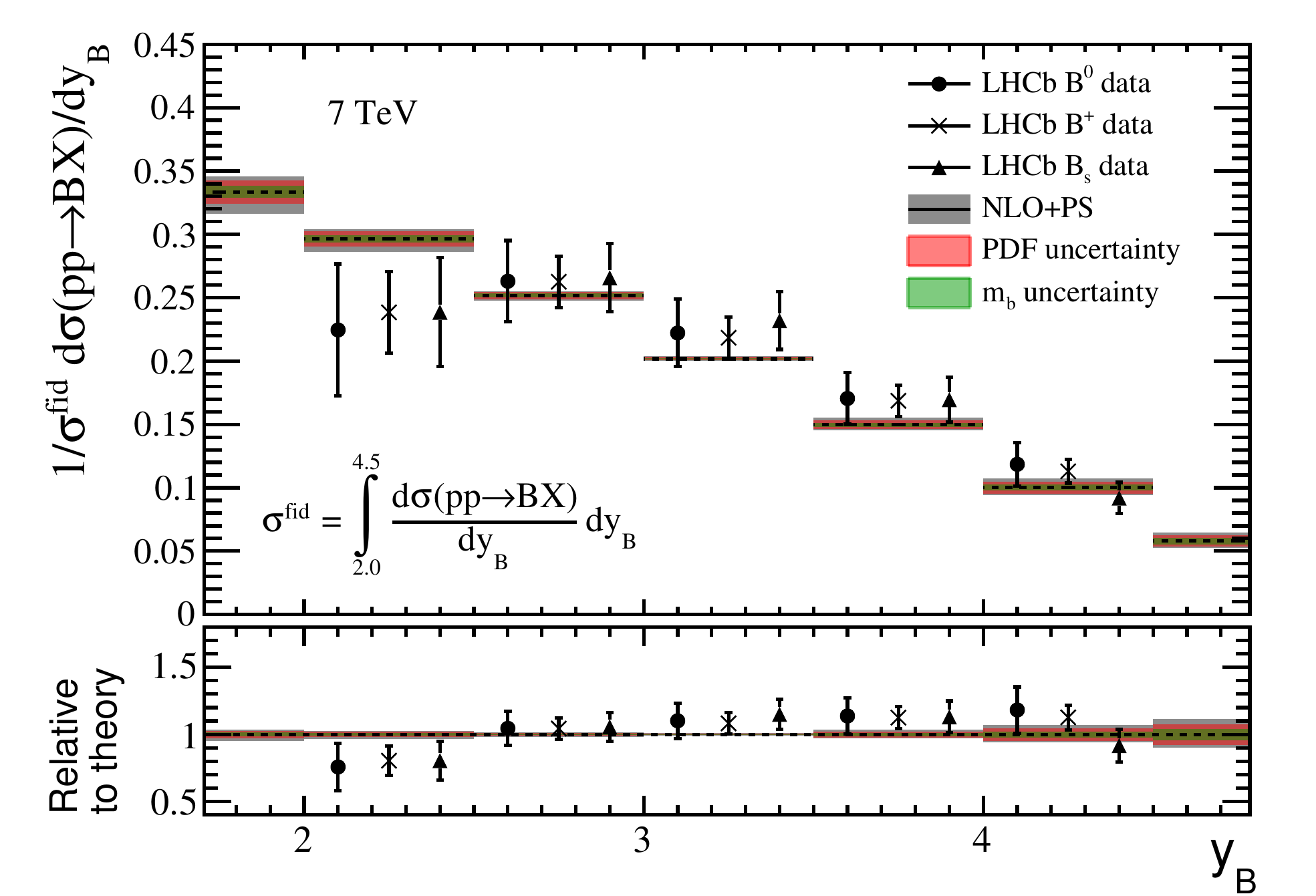}}
\end{center}
\vspace{0mm}
\caption{Absolute (left) and normalised (right) differential LHCb $B$ hadron cross section data at 7~TeV. 
The theoretical uncertainty on the {\sc\small NLO$+$PS} accurate prediction is obtained as the sum in quadrature of the scale, PDF, and $m_b$ uncertainties.}
\label{fig:BtoJpsiX}
\end{figure}

The comparison to data is shown in Fig.~\ref{fig:BtoJpsiX}, where both the
absolute (left) and normalised (right) $B$ hadron rapidity distributions are shown. Again,
in the lower panel both the theoretical predictions and data are shown normalised to the central
theory prediction. In this case, the shown theoretical prediction corresponds to the $B^{+}$ hadron 
final state, where a fragmentation fraction of $f(b\to B_u = 0.337)$ has been applied. Excluding 
the value of the fragmentation fraction, the individual distributions for $B^{0}$ and $B_s$ 
hadrons are extremely similar and are therefore not shown. 
Finally, in the case of the absolute cross for $B_s$ production, the experimental 
data has been multiplied by a normalisation factor of 
$\sigma^{\rm fid.}_{y}(B^{+})/\sigma^{\rm fid.}_{y}(B_s) \approx 3.7$.
This normalisation is applied to allow the $B_s$ cross section to be
compared with the other $B$ hadron final states simultaneously.
As demonstrated by this comparison, the individual measurements of all 
three $B$ hadron are self consistent, and also consistent within uncertainties 
with the theoretical predictions for both the absolute and normalised cross section. 
However, there is some tendency for the data to undershoot the predictions in 
the region $y^B \in [2.0,2.5]$.

\begin{figure}[t!]
\begin{center}
\makebox{\includegraphics[width=0.49\columnwidth]{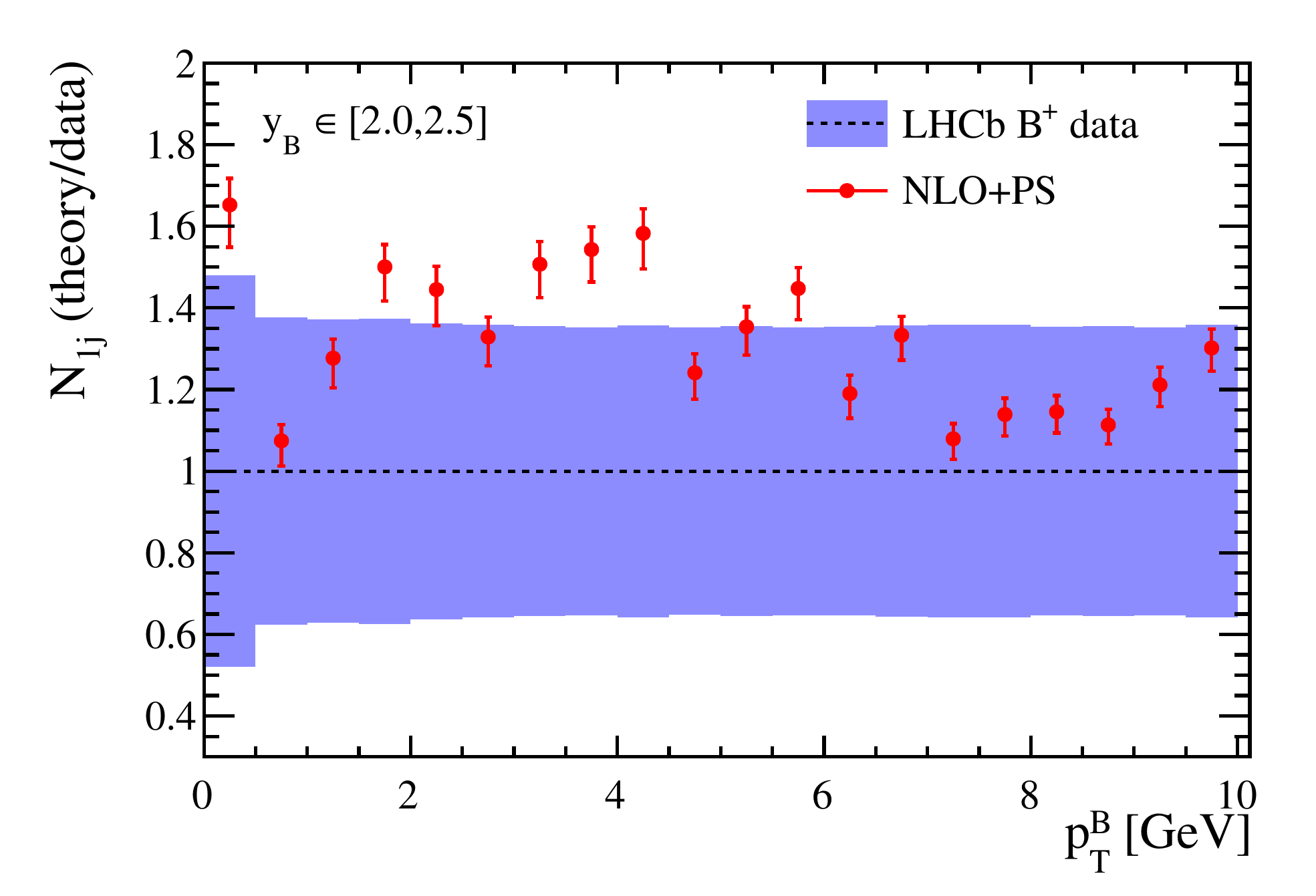}}
\makebox{\includegraphics[width=0.49\columnwidth]{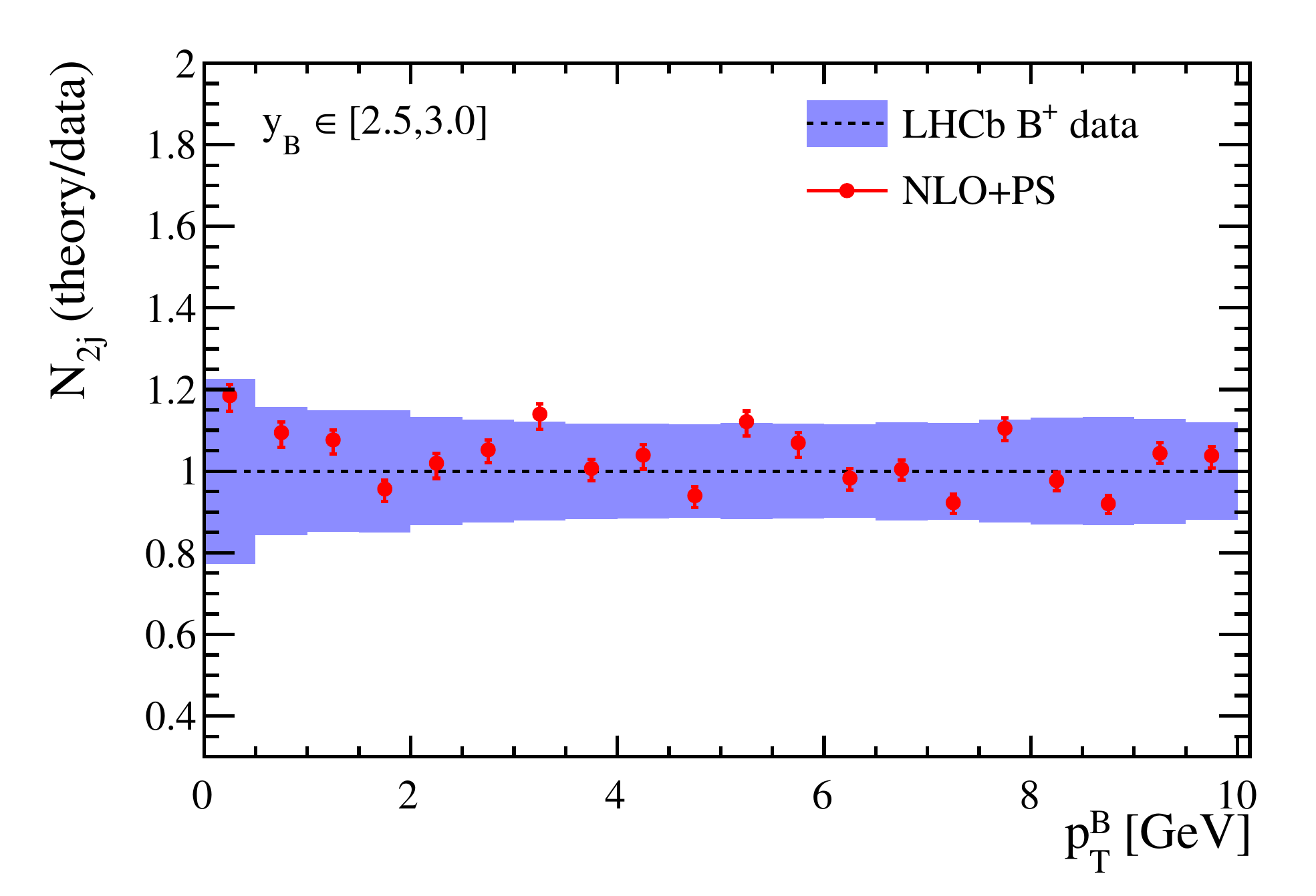}}
\end{center}
\vspace{0mm}
\caption{The normalised double differential $B^+$ cross section for $y_B \in [2.0,2.5]$ (left) and $y_B \in [2.5,3.0]$ (right). For both plots, theory and data are normalised to the central data point in each bin.}
\label{fig:BtoJpsiXpt}
\end{figure}

In addition to the $p_T$ integrated distributions, a similar comparison can also be performed
for the double differential data. This is done by normalising the data (for each $p_T$ bin)
to that in the central rapidity bin $y^B_{\rm ref.} \in [3.0,3.5]$~\cite{Zenaiev:2015rfa}. Therefore, for
a given $y_B$ ($i$) and $p_T^B$ ($j$) bin this observable is defined as
\begin{align} \label{eq:doublediff}
N_{ij} =  \frac{d^2 \sigma(pp\to BX)}{d y^B_i  d (p_T^B)_j} \bigg{/} \frac{d^2 \sigma(pp\to BX)}{d y^B_{\rm ref.}  d (p_T^B)_j} \,.
\end{align}
A comparison of selected data and predictions for this observable are shown in Fig.~\ref{fig:BtoJpsiXpt}, 
where the double differential $B^{+}$ data (which is most precise) is compared with the corresponding 
theoretical predictions.
The lowest rapidity $y_B \in [2.0,2.5]$ region is shown in the left plot, while the neighbouring bin $y_B \in [2.5,3.0]$
is shown in the right. In both cases, the predictions and data are normalised to the central
value of the data in each bin. In the lower rapidity region of $y_B \in [2.0,2.5]$ (and for $p_T^B < 7$~GeV), the data 
tends to systematically lie below the theoretical predictions. Although not shown, this behaviour is also observed for 
$B^{0}$, and $B_s$ hadrons. In contrast, excellent agreement is found (within uncertainties) for the other rapidity
bins, shown in the right plot for $y_B \in [2.5,3.0]$. 
While the tension at low $y_B \in [2.0,2.5]$ is rather mild, it is worth mentioning 
that the experimental uncertainties (which have been constructed) are again likely over estimated 
since only the luminosity and branching ratio uncertainties are treated as correlated. 
The agreement with data could be better quantified with the experimental bin-by-bin correlations.

In summary, the 7~TeV $B\to J/\psi X$ cross section data (both absolute and normalised) are
consistent with the theoretical predictions presented differentially in $p_T^B$ and $y_B$.
There is some tendency for the normalised $B$ hadron data (observed for $B^+$, $B^0$ and $B_s$ final states)
to undershoot the theoretical predictions in the region $y_B \in [2.0,2.5]$ and $p_T^B < 7$~GeV.
This same trend is observed for the pseudorapidity dependent measurement at 7~TeV (but not at 13~TeV).
It will be interesting to see if similar behaviour is observed in a corresponding 
13~TeV measurement. In addition, as proposed in~\cite{Gauld:2015yia,Cacciari:2015fta}, it would 
be useful for the ratio of 13 and 7~TeV cross section measurements to be performed 
(double) differentially in $p_T^B$ (and $y_B$).

\section{Ratio of $B$ hadron cross section data} \label{Ratio}
The general motivation for considering a ratio of cross section measurements
at different CoM energies is that the theoretical (and many experimental) 
uncertainties for a specific process are correlated between different CoM energies. 
Therefore, many sources of uncertainty partially cancel when constructing such a ratio. In some cases,
this results in a dramatic reduction in scale uncertainties allowing sensitivity to PDFs, 
or both experimental and theoretical uncertainties may be reduced to an extent
that these measurements can be used for luminosity determination of searches for
the effects of physics beyond the Standard Model~\cite{Mangano:2012mh}.
As mentioned in the Introduction, this method is particularly
useful when considering $B$ (and $D$) hadron production, as 
this is a process which is otherwise overwhelmed by large scale uncertainties.
At the same time, the rate of the cross section growth with increasing CoM energy
provides information on the shape of the gluon PDF at both small- and large-$x$.

\begin{figure}[t!]
\begin{center}
\makebox{\includegraphics[width=0.49\columnwidth]{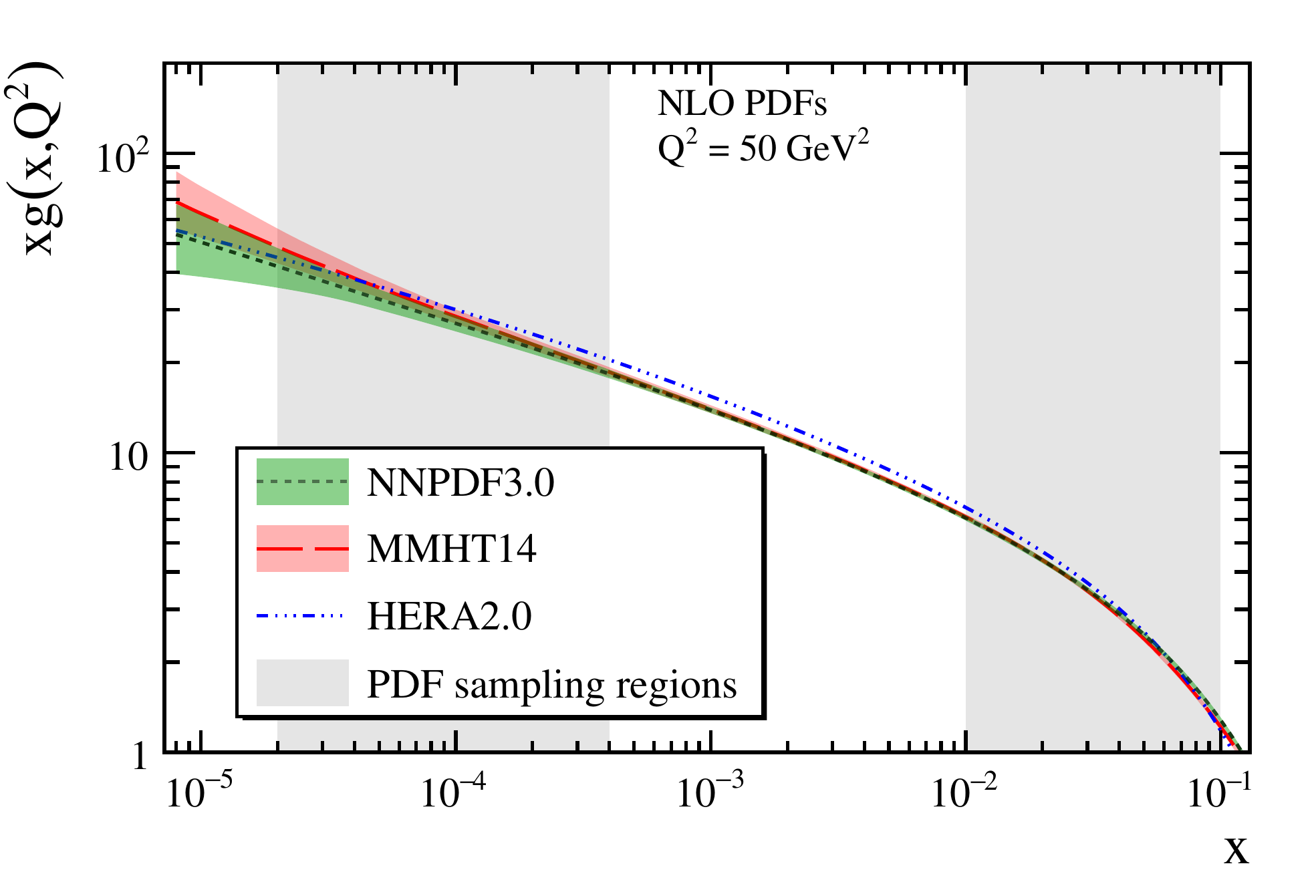}}
\makebox{\includegraphics[width=0.49\columnwidth]{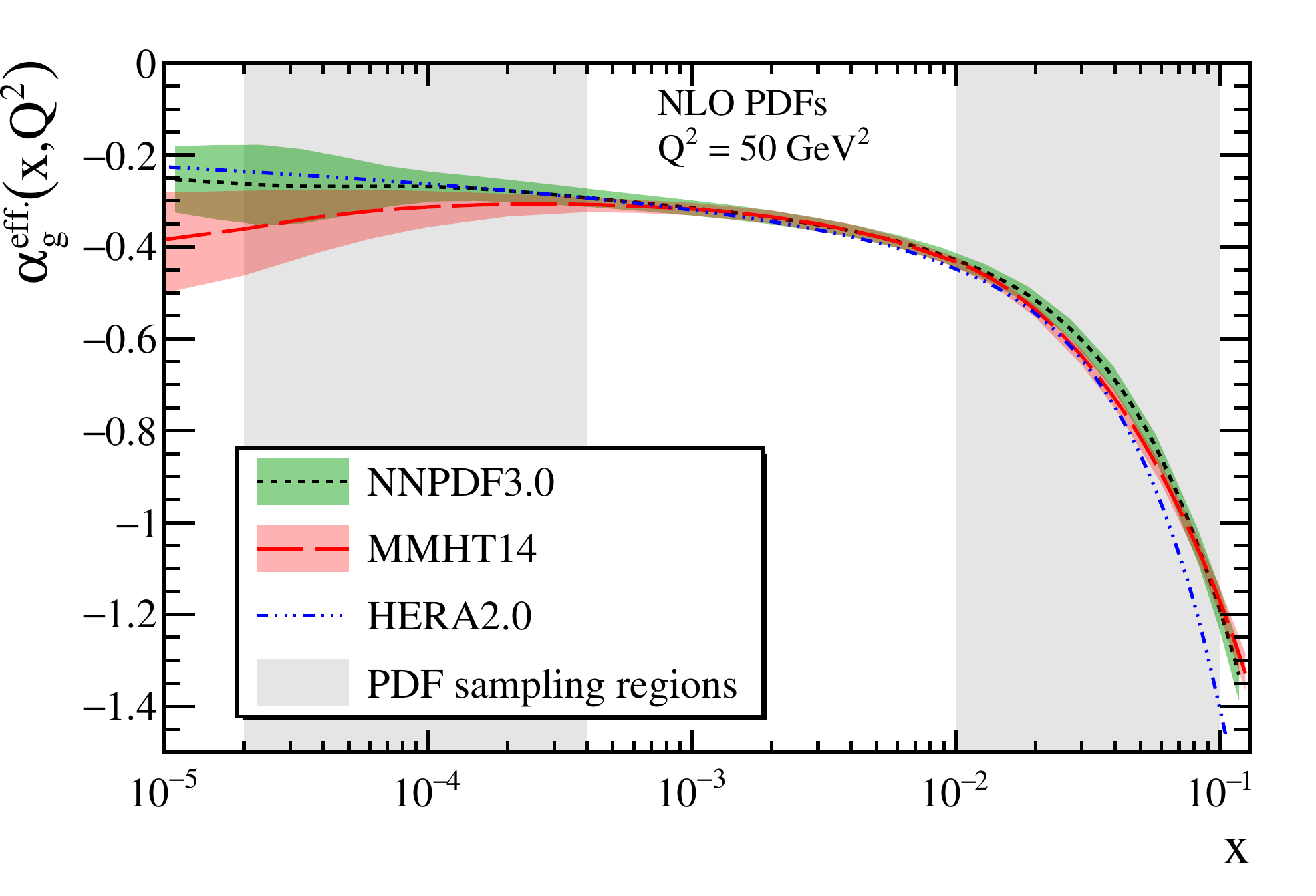}}
\end{center}
\vspace{0mm}
\caption{Left: the evolved gluon PDF $xg$ at the scale $Q^2 = 50$~GeV$^2$. 
Right: the effective gluon exponent $\alpha_{\rm g}^{\rm eff.}$ also at the scale $Q^2 = 50$~GeV$^2$. In both cases,
the approximate PDF sampling regions of forward $B$ hadron production are highlighted.}
\label{fig:alpha}
\end{figure}

To better understand the behaviour of the $B$ hadron ratio data considered in
this Section, it will be useful to introduce the following quantity
\beq \label{eq:alpha}
\alpha_{\rm g}^{\rm eff.}(x,Q^2) = \frac{ \partial \ln \left[ x g (x,Q^2) \right] }{\partial \ln x} \,,
\eeq
which effectively describes the logarithmic growth of the gluon PDF with respect to $x$, and
has recently been used to study the asymptotic behaviour of PDFs~\cite{Ball:2016spl}. 
This is a useful quantity when considering the ratio of $B$ or $D$ hadron production measurements, since
this observable is sensitive to exactly this growth. 
The computation of $\alpha_{\rm g}^{\rm eff.}(x,Q^2)$ for different PDF sets can be performed 
numerically using the LHAPDF interface, for which the PDF sets are provided as data 
files on grids in $x$ and $Q^2$ space. The derivative in Eq.~(\ref{eq:alpha}) can be performed at each $x$ point on
the grid by fitting a polynomial to the values of $\ln \left[xg(x,Q^2)\right]$ obtained for 
the neighbouring grid points in $x$. For the results shown in this work, a polynomial of order 3
is fitted to the central $x$ point and the four neighbouring points in either direction.
The results of this procedure are shown in Fig.~\ref{fig:alpha}, where both the gluon PDF 
(left) and $\alpha_{\rm g}^{\rm eff.}(x,Q^2)$ (right) are shown for the baseline NNPDF3.0 NLO PDF, 
as well as the MMHT14~\cite{Harland-Lang:2015qea} and HERA2.0~\cite{Abramowicz:2015mha} gluon PDFs. 
While not shown here, the effective exponents for the NLO gluon PDF from CJ15~\cite{Accardi:2016qay}, 
ABM11~\cite{Alekhin:2012ig} and CT14~\cite{Dulat:2015mca} PDF fits exhibit the same behaviour as those shown. 
That is, at large-$x$ ($x\sim0.1$) the gluon PDF grows extremely quickly as it is generated by the valence 
PDF content, while at low-$x$ the logarithmic growth becomes approximately constant. As
demonstrated in Fig.~\ref{fig:x1x2}, both large- and small-$x$ regions are important for describing 
the forward $B$ hadron ratio data.
The remainder of this Section will be dedicated to studying various incarnations
of cross section ratios.

\subsection{Fiducial and differential ratio}
Before discussing the differential data, it is instructive to first consider
the ratio of the fiducial cross section measurements. This observable is defined as
\beq \label{eq:Rfid}
R^{\rm fid.}_{13/7} = \frac{ \sigma_{\eta_B}^{\rm fid.}(13{\rm~TeV})} {\sigma_{\eta_B}^{\rm fid.}(7{\rm~TeV})} \,,
\eeq
where the fiducial cross section $\sigma_{\eta_B}^{\rm fid.}$ has previously been defined in Eq.~(\ref{eq:Beta}).
The experimental measurement and corresponding theoretical prediction are provided below
\begin{align} \nonumber
R^{\rm fid.}_{13/7}({\rm LHCb}) &= 2.14\pm{0.02}\,{\rm (stat.)}\pm{0.13}\,{\rm (sys.)} \,, \\[1mm]
R^{\rm fid.}_{13/7}({\rm {\sc\small NLO+PS}}) &= 1.784^{+0.022}_{-0.020}\,{(m_b)}\,^{+0.043}_{-0.043}\,{\rm (PDF)}  \,^{+0.061}_{-0.104}{\rm (Scale)} \,.
\end{align}
In both cases, the breakdown of the various contributions to the uncertainty are provided. 
For the theoretical prediction, the scale uncertainties are still the dominant source of uncertainty, 
since the gluon PDF is predominantly sampled in the region which is constrained to a few $\%$ uncertainty. 
For example, when computing the 7~TeV fiducial cross section at LO, the mean sampling values are 
$\bar{x}_1 \sim 3\cdot 10^{-2}$ and $\bar{x}_2 \sim 2 \cdot 10^{-4}$ at a scale of $\overline{m}_T^2 \sim 50{\rm~GeV}^2$.

\begin{figure}[ht!]
\begin{center}
\makebox{\includegraphics[width=0.49\columnwidth]{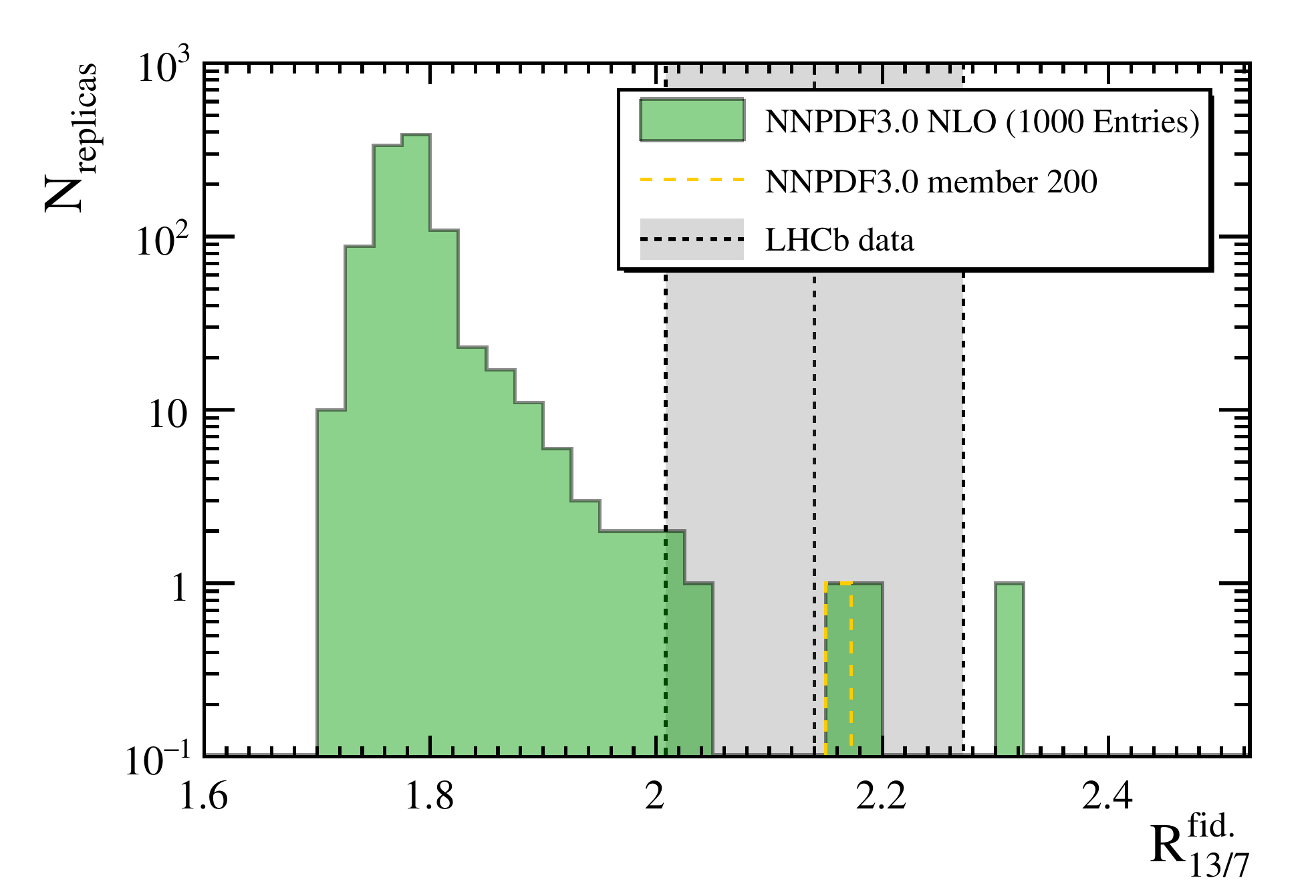}}
\end{center}
\vspace{0mm}
\caption{Predictions for $R^{\rm fid.}_{13/7}$ obtained with the
NNPDF3.0 NLO 1000 PDF replica set. The value of the LHCb measurement is also indicated.}
\label{fig:RfidRepl}
\end{figure}

The data is 2.7$\sigma$ above the central theory prediction, and
the predictions and data are consistent within their 2$\sigma$ CL uncertainties. Although disfavoured
by the baseline PDF set (NNPDF3.0), it is in principle still possible to accommodate this
behaviour with a more steeply rising gluon PDF at low-$x$. This can be seen
in Fig.~\ref{fig:RfidRepl} where the individual replica predictions obtained with the NNPDF3.0 NLO 
1000 PDF replica set are shown. The handful of outliers which are consistent with the LHCb
data have exactly this feature. For example, replica member 200 (which is closest to the data) leads to
a prediction of $R^{\rm fid.}_{13/7} = 2.16$. 
However, it is worth mentioning that in the recent analysis of
the forward LHCb $D$ hadron data~\cite{Gauld:2016kpd}, this same 
replica member provided an extremely poor description of the normalised $D$ hadron
cross section data at 5, 7, and 13~TeV.

\begin{figure}[t!]
\begin{center}
\makebox{\includegraphics[width=0.49\columnwidth]{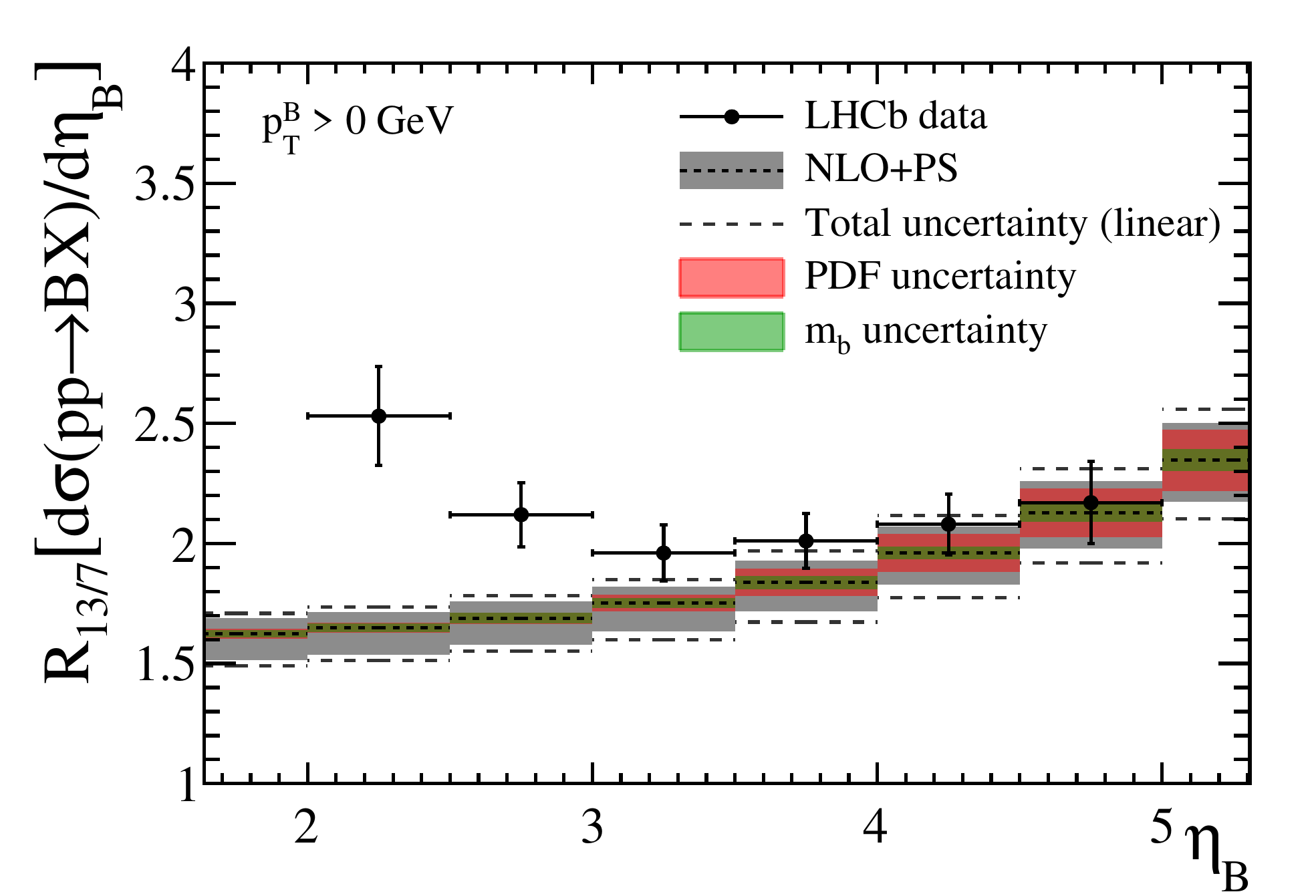}}
\makebox{\includegraphics[width=0.49\columnwidth]{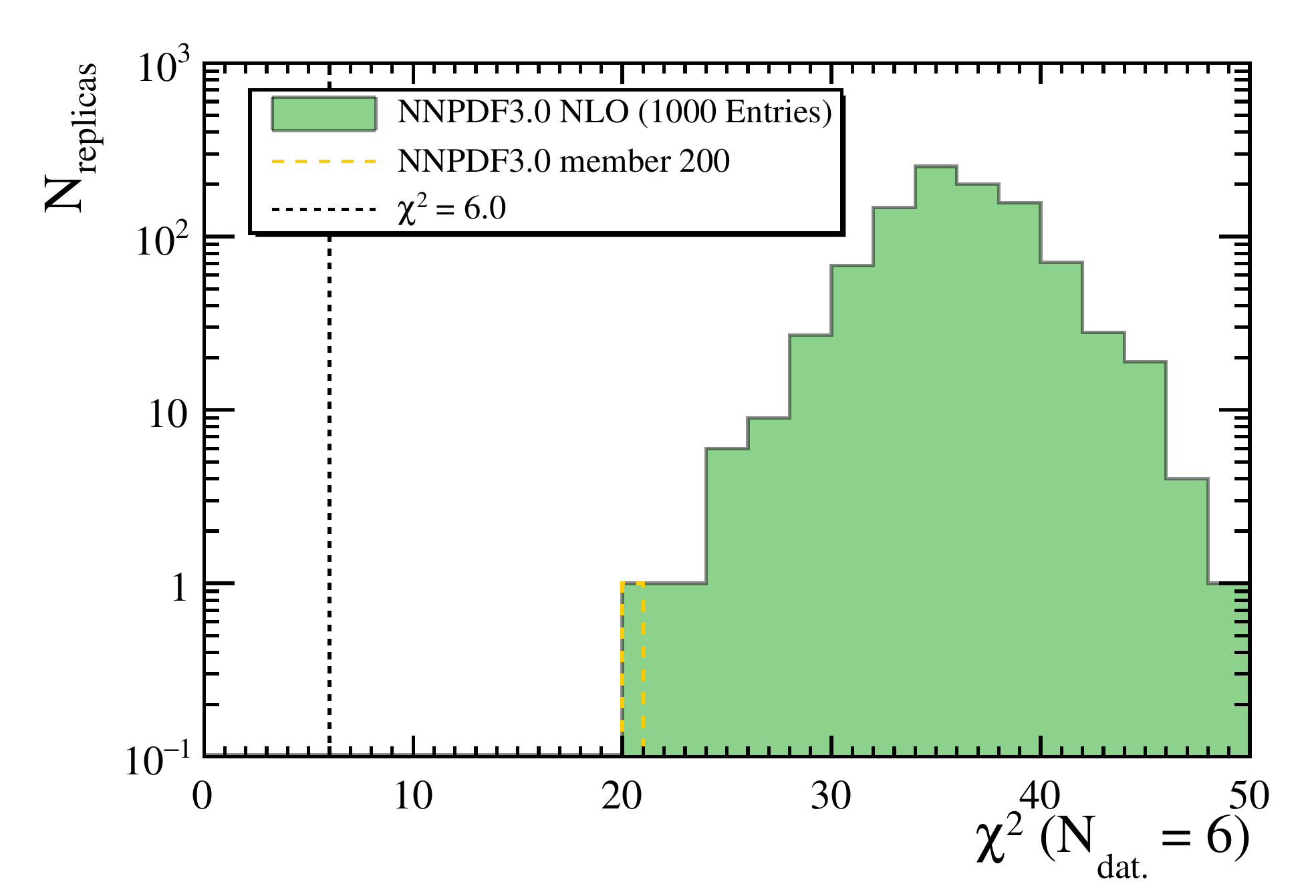}}
\end{center}
\vspace{0mm}
\caption{Left: The differential ratio ($R_{13/7}$) of inclusive $B$ production with respect to pseudorapidity measured within the LHCb acceptance. Right: The $\chi^2$ values obtained for each PDF replica in comparison to the differential ratio data.}
\label{fig:Ratio}
\end{figure}

The cross section ratio measurement by LHCb is also presented
differentially with respect to $\eta_B$, according to 
\beq \label{eq:Rdiff}
R_{13/7} \left[ d\sigma(pp\to BX)/d\eta_B \right]=\frac{d\sigma_{13}(pp\to BX)}{d\eta_B}  \bigg{/} \frac{d\sigma_{7}(pp\to BX)}{d\eta_B} \,.
\eeq
The comparison of the theoretical predictions to data for this observable are provided
in Fig.~\ref{fig:Ratio} (left). In this case, the individual contributions from PDF and
and $m_b$ uncertainties are also shown, and the more conservative `linear' combination
of uncertainties is provided. In the lower pseudorapidity region of 
$\eta_B \in [2.0,3.0]$ the scale uncertainties are dominant, and the PDF uncertainties
become more significant at high pseudorapidity as the gluon PDF is probed at smaller
values of $x$ (see Fig.~\ref{fig:x1x2}, right).
The behaviour of the theoretical prediction is also easy to understand by examining
Fig.~\ref{fig:alpha}. For increasing $\eta_B$ values, the ratio becomes more sensitive
to the gluon PDF at larger (smaller) $x_1$ ($x_2$) values. In the low-$x$ region, the logarithmic
growth of the gluon PDF is approximately flat which results in an approximately pseudorapidity independent
contribution to the ratio. At larger $x$ values, the growth of the gluon PDF accelerates 
with increasing $x$, which results in a larger contribution to the ratio with increasing $\eta_B$.

However, this behaviour is clearly not observed in the LHCb data where
the ratio is largest in lowest pseudorapidity region of $\eta_B \in [2.0,3.0]$.
In fact, the first data point ($\eta_B \in [2.0,2.5]$) is 4.3$\sigma$ 
above the central theory prediction, and 4.0$\sigma$ 
with respect to the conservative upper theoretical uncertainty.
The overall agreement with the data is extremely poor and, unlike the fiducial cross section,
there are no individual replica PDF members which provide an adequate description of the
data. This is shown in Fig.~\ref{fig:Ratio} (right), where the $\chi^2/{\rm N_{dat}}$ of the differential
ratio data is computed with respect to each of the 1000 replica PDF members. The mean value is 
$\overline{\chi}^2/{\rm N_{dat}} = 36/6$, and the minimum value (member 200) is $\chi^2/{\rm N_{dat}} = 21/6$.

\begin{figure}[ht!]
\begin{center}
\makebox{\includegraphics[width=0.49\columnwidth]{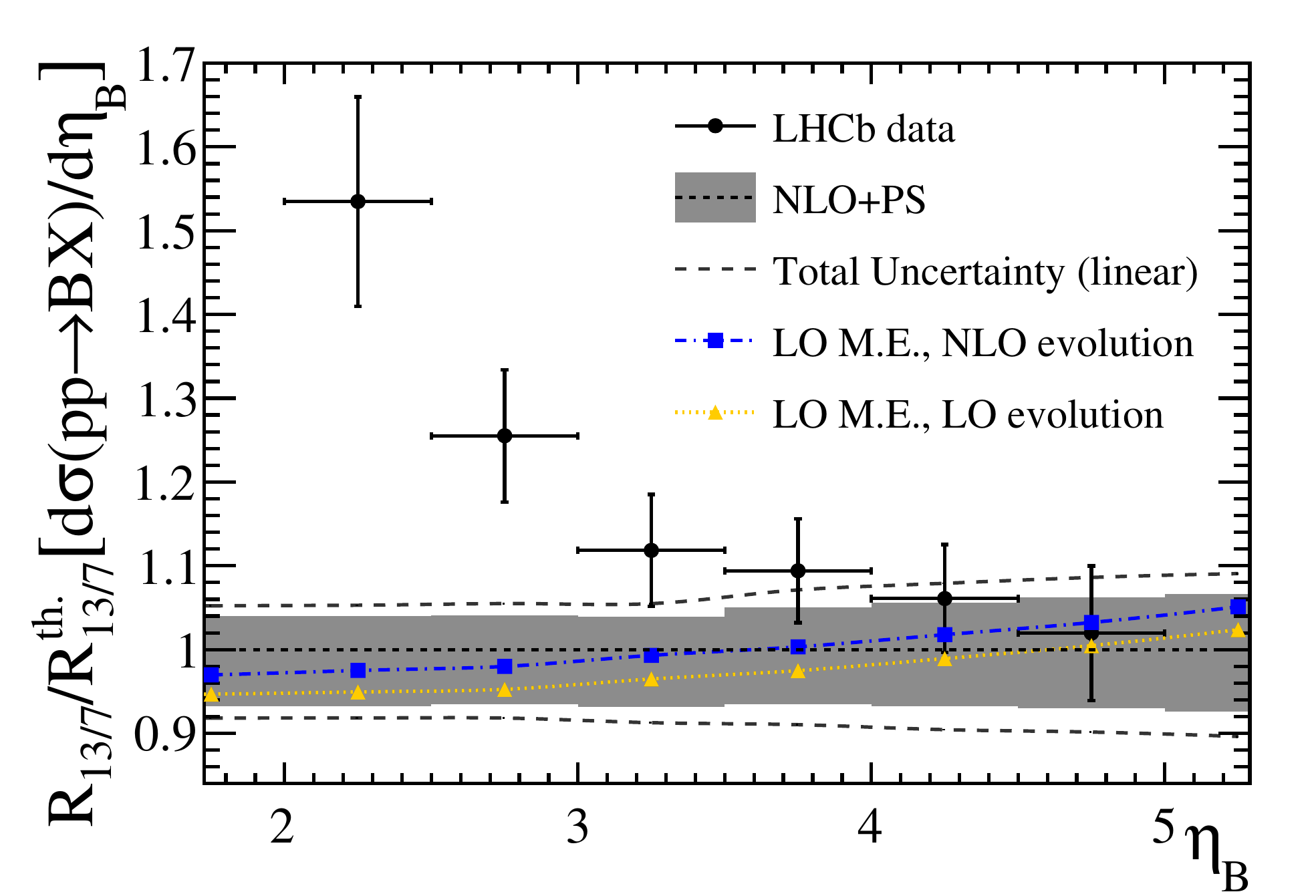}}
\end{center}
\vspace{0mm}
\caption{The differential ratio ($R_{13/7}$) normalised to the central theory prediction, 
including predictions obtained with LO matrix elements and (N)LO PDF 
and $\alpha_s$ evolution.}
\label{fig:Perturb}
\end{figure}

As a cross check of the theoretical predictions, it is important to study the perturbative
stability of the ratio observable defined in Eq.~(\ref{eq:Rdiff}). This is shown in Fig.~\ref{fig:Perturb}, 
where both data and theoretical predictions are shown normalised to the central theory prediction. 
In this case, predictions are shown when LO matrix elements (M.E.) are used for the evaluation of the partonic cross section,
which is then convoluted with the baseline PDFs (and $\alpha_s$) evolved at either LO or NLO accuracy ---
the evolution is performed with the {\sc\small APFEL} PDF evolution libraries~\cite{Bertone:2013vaa}.
This exercise demonstrates that the perturbative corrections, both through the evolution and the partonic 
cross section, are mild (each below 4\%). It is therefore unexpected that NNLO QCD corrections would
dramatically alter the theoretical predictions for this ratio observable. 
%
Another source of uncertainty not included in the total uncertainty is related to the
treatment of the heavy quark fragmentation. The potential impact of this uncertainty 
has been assessed by varying the Lund-Bowler $b$-quark fragmentation variable $r_b$ 
within the range of $r_b \in [0.67,1.00]$, and by additionally showering events with the 
non-default {\sc\small Pythia8} Tune 4C. Further to this, the {\sc\small POWHEG} events 
have also been showered with the {\sc\small Herwig7.0} {\sc\small PS}~\cite{Bahr:2008pv,Bellm:2015jjp}\footnote{I am grateful to P. Schichtel and J. Bellm for assistance using {\sc\small Herwig7.0}.}. In all cases, the resultant ratio predictions
differ from the central prediction by less than 3\% within the region of $\eta_B \in [2.0,5.0]$,
which justifies not including this contribution in the total uncertainty.

To understand the origin of the tension observed in data, it will be useful to define kinematically
shifted ratio observables which will be considered in the remainder of this Section.

\subsection{Rapidity shifted differential ratio}

\begin{figure}[t!]
\begin{center}
\makebox{\includegraphics[width=0.49\columnwidth]{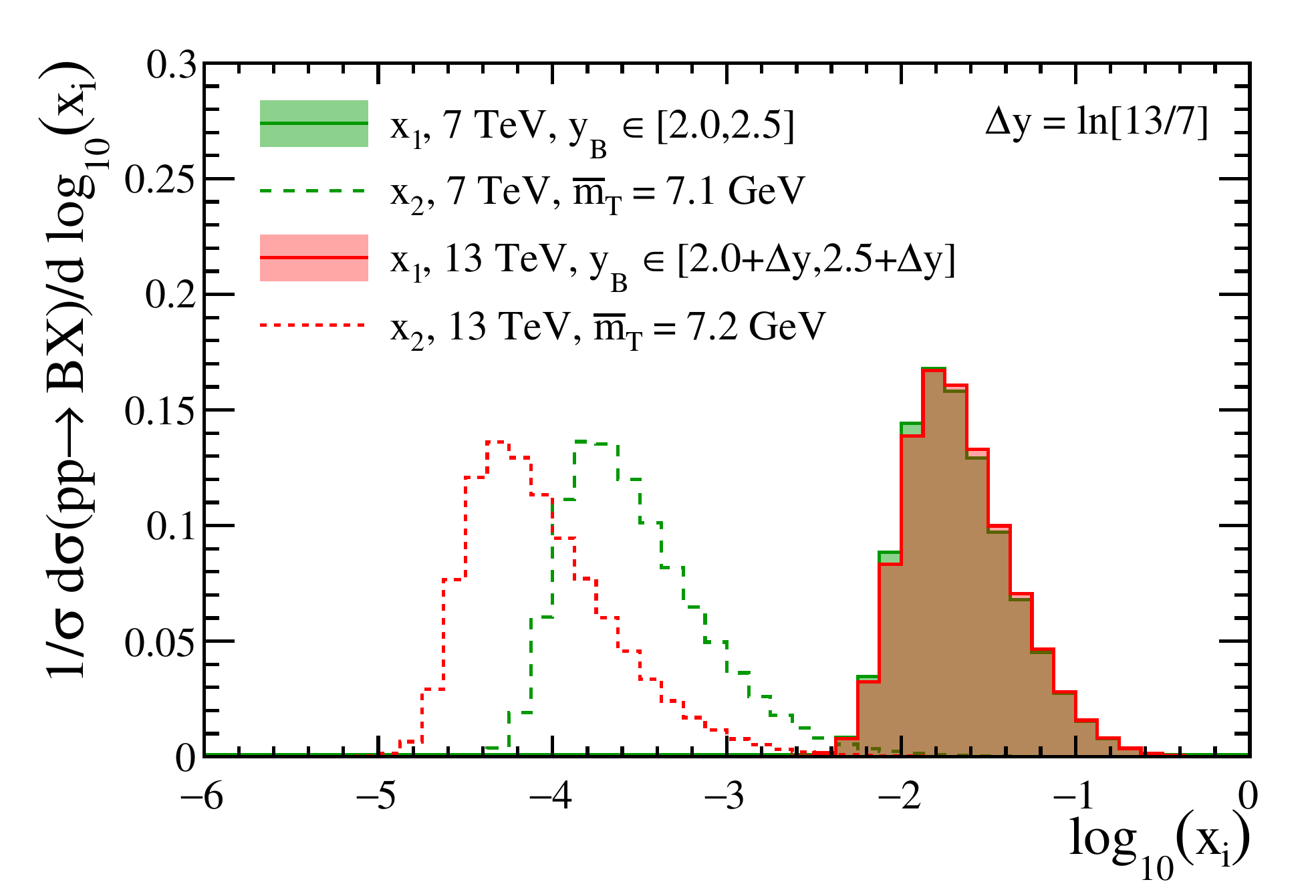}}
\makebox{\includegraphics[width=0.49\columnwidth]{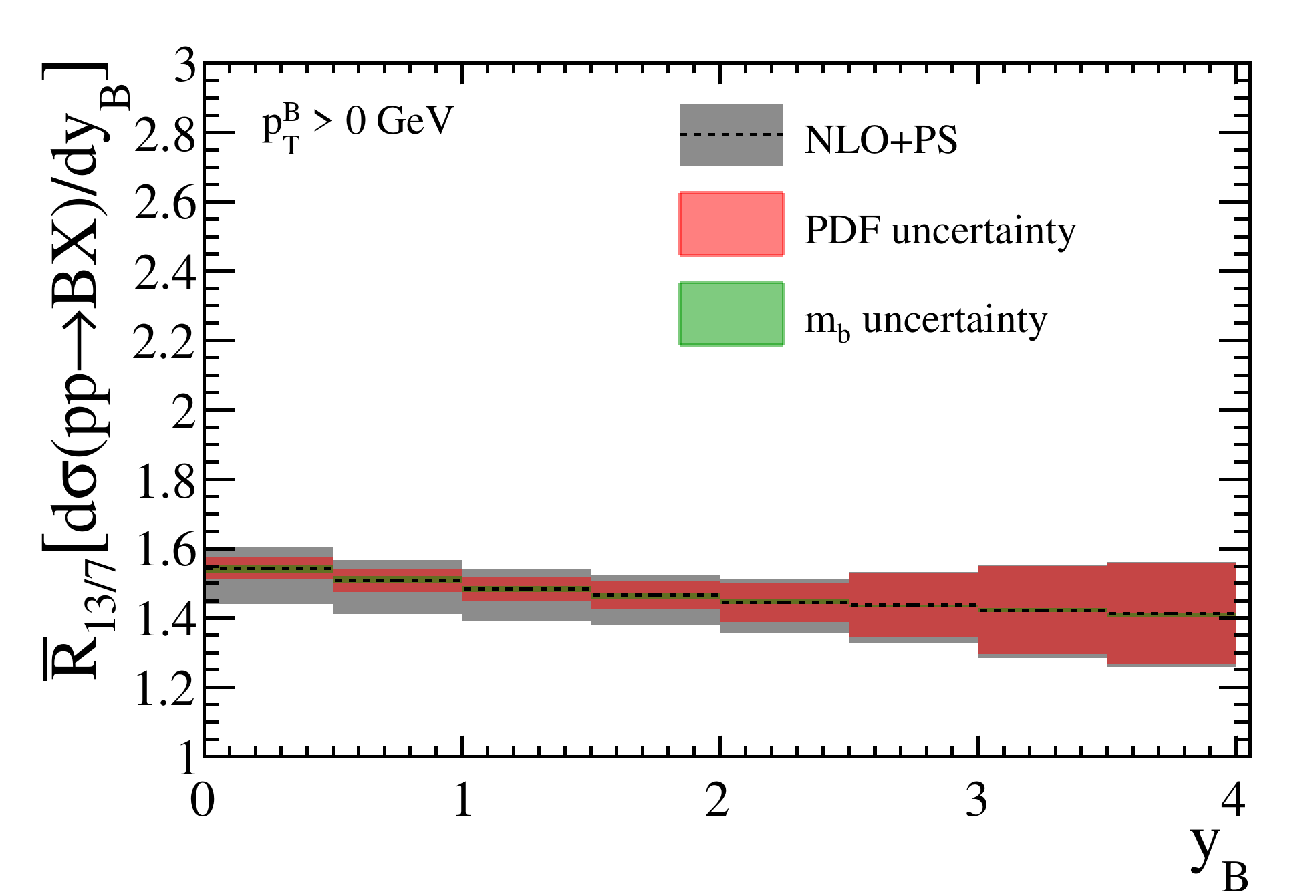}}
\end{center}
\vspace{0mm}
\caption{Left: The LO $B$ hadron cross section as a function of the $x_{1,2}$ within the LHCb acceptance
for specific $B$ hadron rapidity regions at 7 and 13~TeV. 
Right: NLO predictions for the ratio $\overline{R}_{13/7}$ of inclusive $B$ production with respect to $y_B$.}
\label{fig:xshift}
\end{figure}

As discussed in the previous Subsection, the behaviour of the differential ratio 
defined in Eq.~(\ref{eq:Rdiff}) depends on both the behaviour of the gluon PDF
at small- and large-$x$ values. This is because increasing $\sqrt{S}$ results in
a shift of the PDF sampling in both $x$ regions according to $\bar{x}^{13}_{1,2} \sim (7/13) \bar{x}^{7}_{1,2}$.
It is however possible to construct a ratio where either the small or large-$x$
PDF regions are aligned. This can be achieved by introducing a rapidity
shift between the kinematic region for which the numerator and denominator
of the ratio are evaluated at.
In heavy quark pair production, the PDF sampling depends on the outgoing rapidities
of both heavy quarks --- see Eq.~(\ref{eq:xsampling}). However, for a given value
of the $b$ quark rapidity $y_b$, the $\bar{b}$ quark rapidity is symmetrically\footnote{Beyond
LO this is not strictly true~\cite{Aaij:2014ywa,Murphy:2015cha,Gauld:2015qha}.
However, since $B$ hadron production at low-$p_T$ is entirely dominated by the symmetric
gluon-fusion initial state, such an asymmetry is not observable.} distributed 
around $y_b$ such that on average $y_{\bar{b}} = y_b$. Therefore, an alignment
of the mean $x$ sampling regions can be achieved by introducing the shift
\begin{align}
\Delta y = \ln \left[ \frac{13 {\rm~TeV} }{ 7 {\rm~TeV} } \right]  = 0.62 \,.
\end{align}
With this shift, one can specifically align (separate) the  $\bar{x}_1$ ($\bar{x}_2$) sampling regions
by introducing the observable
\begin{align} \label{eq:RbarY}
\overline{R}_{13/7} \left[ d\sigma(pp\to BX)/dy_B \right]=\frac{d\sigma_{13}(pp\to BX)}{dy_B^{\prime}}  \bigg{/} \frac{d\sigma_{7}(pp\to BX)}{dy_B} \,,
\end{align}
where the rapidity shift is introduced in the numerator through $dy_B^{\prime} = dy_B + \Delta y$.
An example of this alignment is shown in Fig.~\ref{fig:xshift} (left), where the 
LO $B$ hadron cross section at 7~TeV is shown as a function of $x_{1,2}$, integrated within
the region of $y_B \in [2.0,2.5]$. The same cross section is shown at 13~TeV with
the shifted integration region of $y_B^{\prime} \in [2.0+\Delta y,2.5+\Delta y]$, demonstrating
the alignment of the large-$x$ regions.
The benefits of introducing this shifted ratio are that the dependence on large-$x$ region is 
eliminated in favour of sensitivity to the low-$x$ region, since the low-$x$ sampling regions 
are separated by a factor of $\bar{x}_2^{13} \approx (7/13)^2 \bar{x}_2^{7}$. At the same time, the 
theoretical uncertainties due to scale and $m_b$ variation are also reduced, since very similar values 
of $\overline{m}_T$ are probed when evaluating the partonic cross section.
This can be seen by examining the {\sc\small NLO$+$PS} accurate predictions for the observable $\overline{R}_{13/7}$,
which are provided in Fig.~\ref{fig:xshift} (right). Although no data is currently available for this ratio 
(which requires the shifted kinematics), future analyses of the LHCb data would have access to this 
observable in the region of $y_{B} \in [2.0,4.0]$. Such a measurement would be very useful for 
understanding the tension observed in the $\eta_B$ dependent measurement.

\begin{figure}[t!]
\begin{center}
\makebox{\includegraphics[width=0.49\columnwidth]{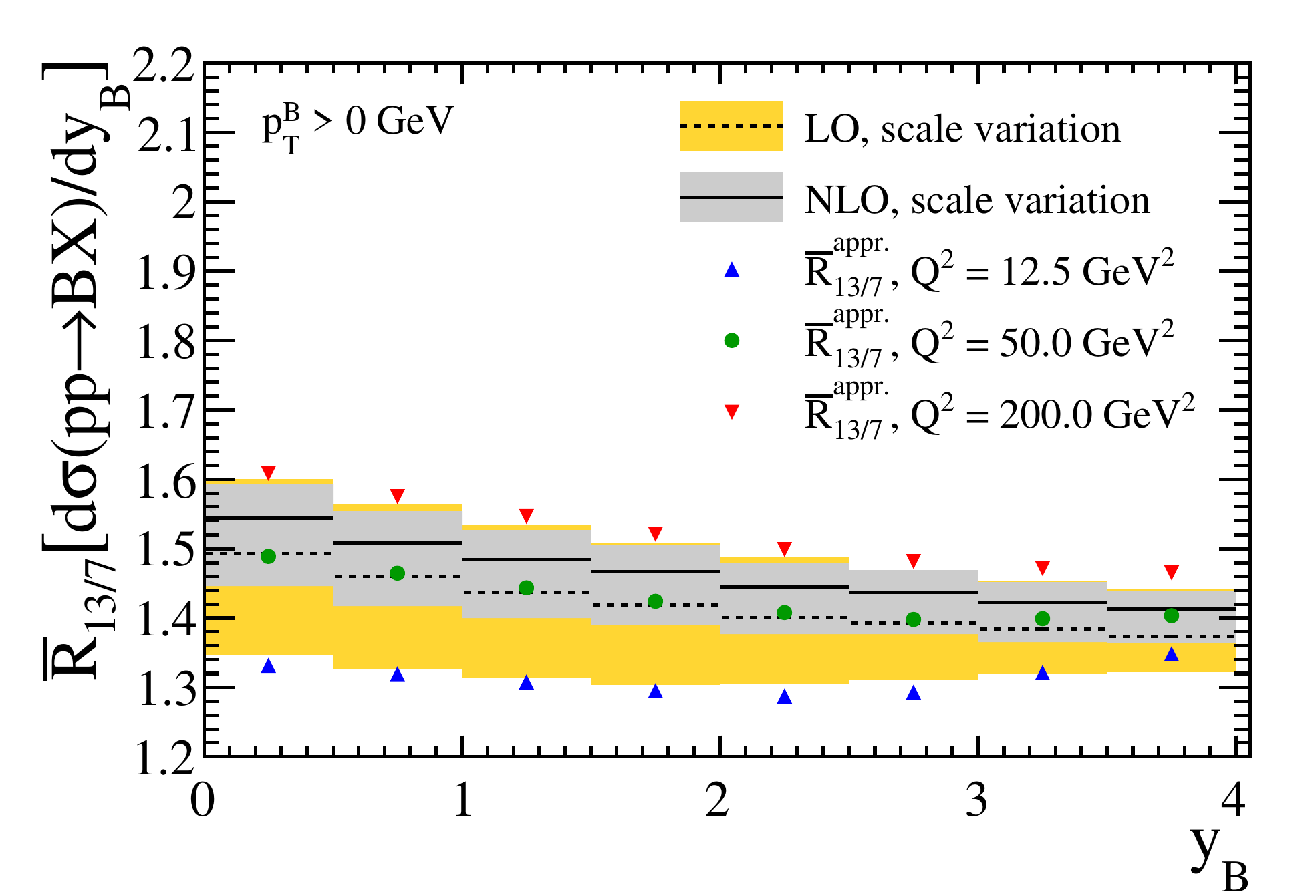}}
\makebox{\includegraphics[width=0.49\columnwidth]{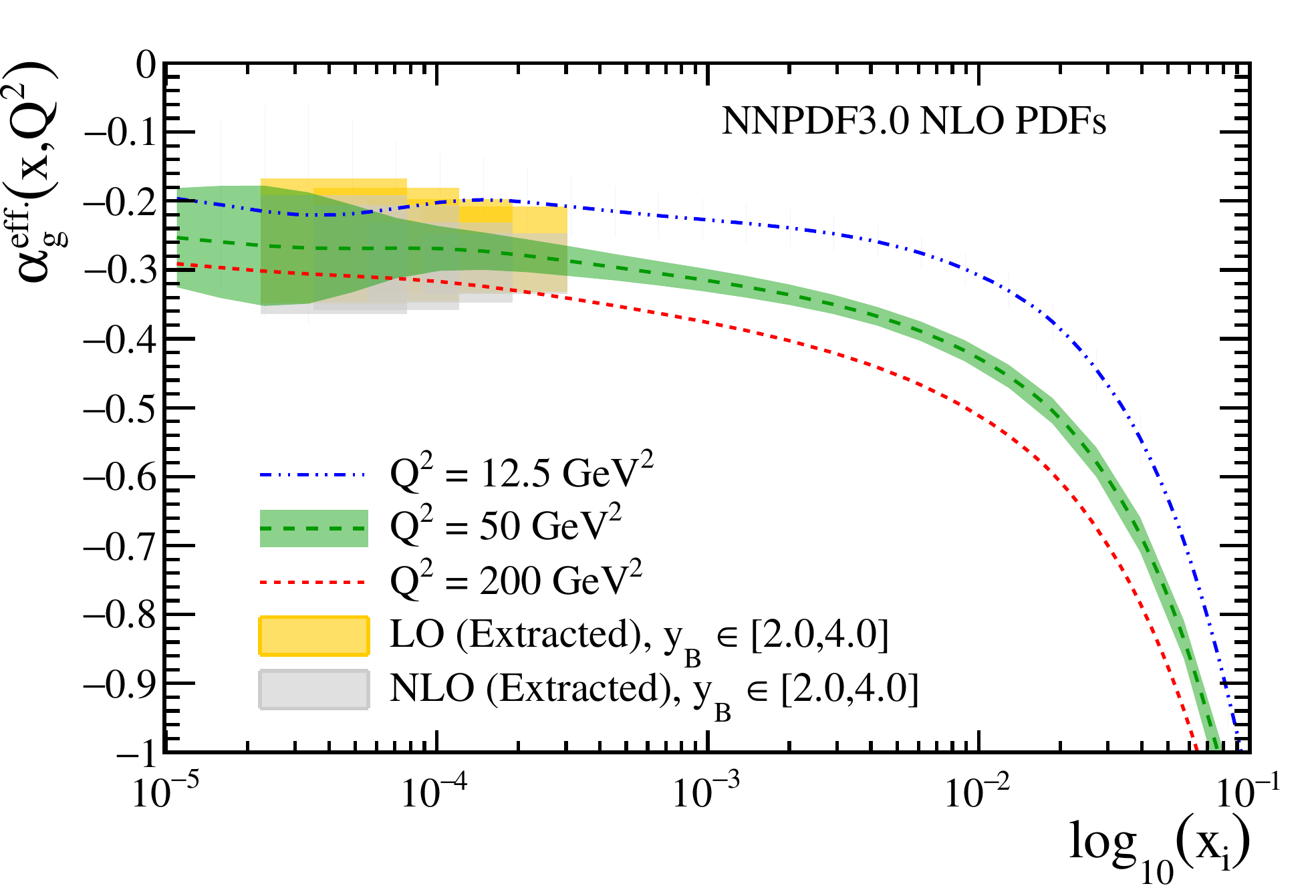}}
\end{center}
\vspace{0mm}
\caption{Left: LO and NLO Predictions for the observable $\overline{R}_{13/7}$, including the effect of scale variation.
In addition, predictions for the quantity $\overline{R}^{\rm appr.}_{13/7}$ for the scale choices of 
$Q^2 = 12.5, 50.0, 200.0{\rm~GeV}^2$ are also shown.
Right: The extracted values of $\alpha_{\rm g}^{\rm eff.}$ obtained from the (N)LO predictions, 
compared to those obtained directly from the input PDFs.
}
\label{fig:RbarApprox}
\end{figure}

An important feature of the shifted ratio observable is that the partonic kinematics which
enter the evaluation of the partonic cross section become highly aligned. Consequently, 
the kinematic dependence of the ratio on partonic cross section is extremely mild, 
and this observable is essentially only sensitive to the growth of the low-$x$ gluon PDF.
This can be demonstrated by studying the variable
\beq \label{eq:RbarYAppr}
\frac{d\overline{R}^{\rm appr.}_{13/7}}{d y_B} = \frac{xg[\bar{x}^{\rm 13~TeV}_2(y_{B}^{\prime}), Q^2]}{xg[\bar{x}^{\rm 7~TeV}_2(y_{B}), Q^2]}\,,
\eeq
where $\bar{x}^{\rm 13~TeV}_2(y_{B}^{\prime})$ and $\bar{x}^{\rm 7~TeV}_2(y_{B})$ correspond
to the mean values of $x_2$ which are sampled when evaluating the $B$ hadron cross section at a given
value of $y_{B}^{\prime}$ at 13~TeV and $y_{B}$ at 7~TeV respectively.
Predictions for this quantity are shown in Fig.~\ref{fig:RbarApprox} (left), along side the actual {\sc\small (N)LO$+$PS}
accurate predictions (both obtained with the baseline NLO PDFs). The prediction of $R^{\rm appr.}_{13/7}$ is performed
by first generating $\bar{x}_2$ values corresponding to rapidity steps of 0.5 from the input value of 
$\bar{x}_2^{\rm 7~TeV}(y_B = 2.25) = 3.0\cdot10^{-4}$ --- this is the mean value of the 
green-dashed distribution in Fig.~\ref{fig:xshift} (left). Explicitly, 
\begin{align} \label{eq:xdep} \nonumber
\bar{x}_2^{\rm 7~TeV}(y_B) &= \bar{x}_2^{\rm 7~TeV}(y_B = 2.25)\cdot e^{2.25-y_B} \,,\\
\bar{x}_2^{\rm 13~TeV}(y_B) &=  (7/13)^2 \bar{x}_2^{\rm 7~TeV}(y_B = 2.25)\cdot  e^{2.25-y_B}\,.
\end{align}
The $\overline{R}^{\rm appr.}_{13/7}$ distribution can then be evaluated numerically with calls to 
the PDF at each of the generated $\bar{x}_2$ values, and is computed for the scale choices 
$Q^2 = 12.5, 50.0, 200.0{\rm~GeV}^2$. These choices correspond to varying the factorisation scale 
by a factor of two around $\mu_f \approx \overline{m}_T \approx 7.1$~GeV. The excellent 
agreement found for the LO prediction and this approximation demonstrate that the shifted
ratio is indeed directly sensitive to the growth of the low-$x$ gluon PDF. Beyond LO, the 
dependence on the choice of the unphysical scale which enters the evaluation of the PDFs
is evidently reduced by the mass factorisation terms present in the partonic cross section.

Of course, a comparison to data for the observable $\overline{R}_{13/7}$ should be performed 
with respect to the most theoretically precise predictions (currently NLO), where the 
dependence on the choice of the unphysical scale is minimal.
However, the reason for introducing the approximate relation $\overline{R}_{13/7}^{\rm appr.}$ is to
first demonstrate that $\overline{R}_{13/7}$ is indeed directly sensitive to the growth of the low-$x$ gluon PDF, 
but to also allow a qualitative study of the behaviour of the $\overline{R}_{13/7}$ observable 
in terms of the quantity $\alpha_{\rm g}^{\rm eff.}(Q^2,x)$. 
The point is that $\overline{R}^{\rm appr.}_{13/7}(y_{B})$ 
measures the growth of the gluon PDF across a given $x$ range 
of $x\in[\bar{x}^{\rm 7~TeV}_2(y_{B}),\bar{x}^{\rm 13~TeV}_2(y_{B}^{\prime})]$.
This is akin to the quantity $\alpha_{\rm g}^{\rm eff.}(Q^2,x)$ which was introduced in 
Eq.~(\ref{eq:alpha}) to study the logarithmic growth of the gluon PDF with respect to $x$,
meaning that it is possible to approximately extract $\alpha_{\rm g}^{\rm eff.}$ from
a differential measurement of $\overline{R}_{13/7}(y_B)$. A 
relation between the two can be obtained according to 
\begin{align} \label{eq:Extract} \nonumber
\frac{d\alpha_{\rm g}^{\rm eff.}}{d y_B} &= \frac{ \ln[ d\overline{R}^{\rm appr.}_{13/7}/d y_B] }{ \ln \left[ \bar{x}^{\rm 13~TeV}_2(y_{B}^{\prime}) / \bar{x}^{\rm 7~TeV}_2(y_{B}) \right] } \,,\\
	&\approx \frac{ \ln[ d\overline{R}_{13/7} /d y_B] }{ \ln \left[ (7/13)^2 \right] } \,,
\end{align}
where the $x$ dependence of $\alpha_{\rm g}^{\rm eff.}$ can be reconstructed in 
a similar fashion to what was done for the $\overline{R}^{\rm appr.}_{13/7}$ predictions
--- see Eq.~(\ref{eq:xdep}).

This approximate relation has been applied to the (N)LO predictions of 
$\overline{R}_{13/7}(y_B)$ (including the total uncertainties)
to extract $\alpha_{\rm g}^{\rm eff.}$ in four experimentally accessible rapidity 
bins within the region of $y_{B} \in [2.0,4.0]$. For each extracted
bin, the upper and lower $x$ values are taken as $\bar{x}^{\rm 7~TeV}_2(y_{B})$
and $\bar{x}^{\rm 13~TeV}_2(y_{B}^{\prime})$. The results of this extraction are 
shown in Fig.~\ref{fig:RbarApprox} (right), and are compared to the quantity 
$\alpha_{\rm g}^{\rm eff.}(x,Q^2)$ obtained directly from the PDFs and computed for the
scale choices $Q^2 = 12.5, 50.0, 200.0{\rm~GeV}^2$.
This method clearly allows the qualitative behaviour of $\alpha_{\rm g}^{\rm eff.}$ to be
extracted from a measurement of $\overline{R}_{13/7}(y_B)$.

\subsection{Pseudorapidity shifted differential ratio}
\begin{figure}[t!]
\begin{center}
\makebox{\includegraphics[width=0.49\columnwidth]{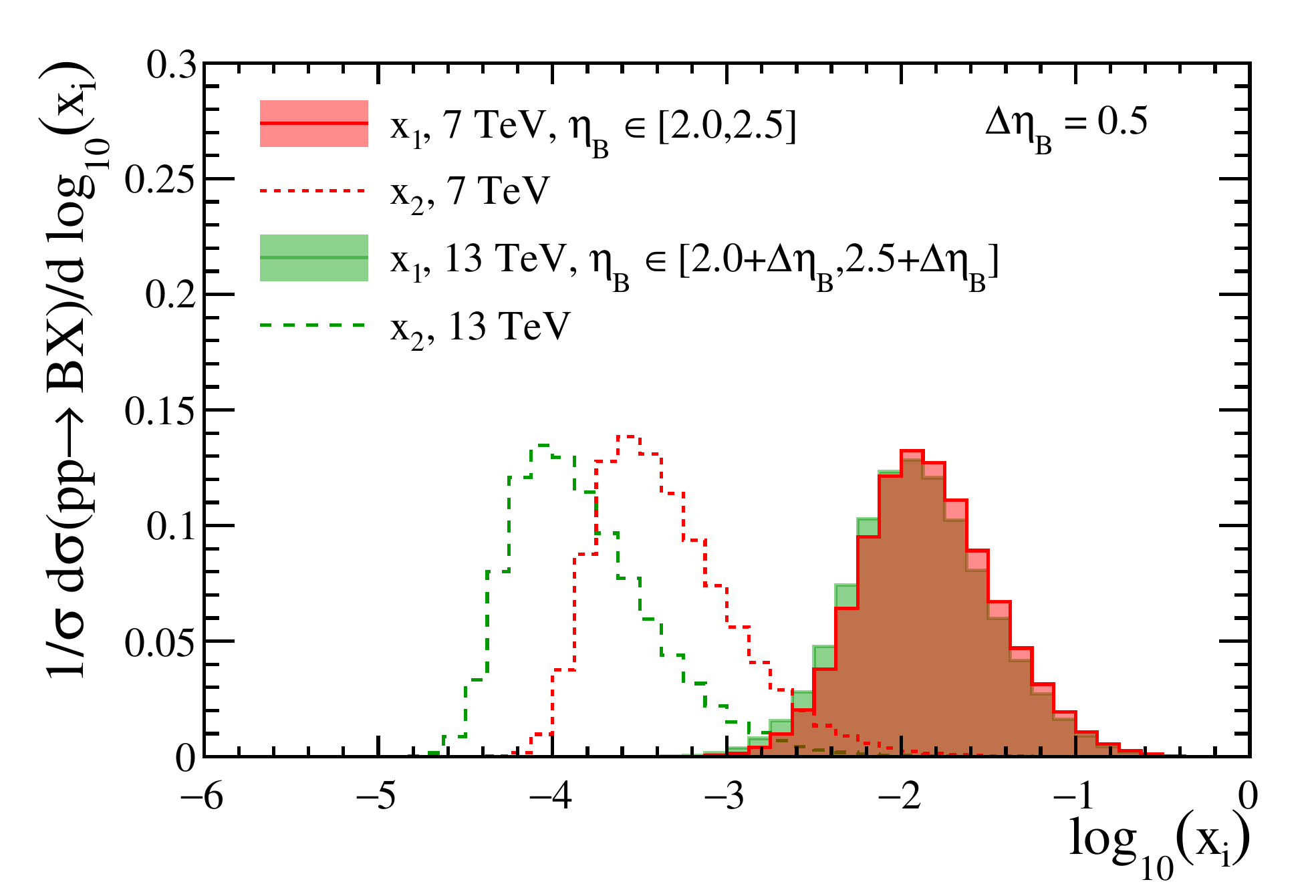}}
\end{center}
\vspace{0mm}
\caption{Same as Fig~\ref{fig:xshift} (left), now with respect to pseudorapidity and with the shift $\Delta \eta_B = 0.5$.}
\label{fig:xshiftEta}
\end{figure}

The LHCb measurement~\cite{Aaij:2016avz} is performed differentially in
pseudorapidity bins of width $\Delta \eta_B = 0.5$, and it is therefore not possible 
to perform the alignment of the $x$ regions as discussed above. However,
a partial alignment can be performed by constructing a pseudorapidity 
shifted ratio according to
\beq \label{eq:Rdiff}
\overline{R}_{13/7} \left[ d\sigma(pp\to BX)/d\eta_B \right]=\frac{d\sigma_{13}(pp\to BX)}{d\eta_B^{\prime}}  \bigg{/} \frac{d\sigma_{7}(pp\to BX)}{d\eta_B} \,, 
\quad d\eta_B^{\prime} = d\eta_B+\Delta \eta_B \,.
\eeq
The success of this partial alignment is shown in Fig.~\ref{fig:xshiftEta}, where the
LO $B$ hadron cross section is shown with respect to $x_{1,2}$, integrated within
the region of $\eta_B \in [2.0,2.5]$. The same cross section is shown at 13~TeV with
the shifted integration region of $\eta_B^{\prime} \in [2.0+\Delta \eta_B,2.5+\Delta \eta_B]$.
The mis-match in the $x_1$ PDF sampling region is approximately $\bar{x}_1^{13}(\eta_B^{\prime}) = 0.9 \bar{x}_1^{7}(\eta_B)$,
and results in the shifted pseudorapidity ratio having minor dependence on the behaviour of the large-$x$ gluon. 
In the region of $\eta_B \in [2.0,4.0]$, this mis-match is estimated to account for a flat correction factor to the ratio of 1.05. 
This `correction factor' is obtained at LO by computing the values $\bar{x}_1^{13}(\eta_B^{\prime})$ and $\bar{x}_1^{7}(\eta_B)$ for each pseudorapidity bin, and by then evaluating $xg\left[\bar{x}_1^{13}(\eta_B^{\prime})\right]/xg\left[\bar{x}_1^{7}(\eta_B)\right]$.
With this exception, the behaviour of this ratio (like the rapidity shifted ratio) is driven by the growth logarithmic growth 
of the gluon at low-$x$ which is approximately flat below $x \sim 10^{-3}$ --- see Fig~\ref{fig:alpha} (right).

\begin{figure}[ht!]
\begin{center}
\makebox{\includegraphics[width=0.49\columnwidth]{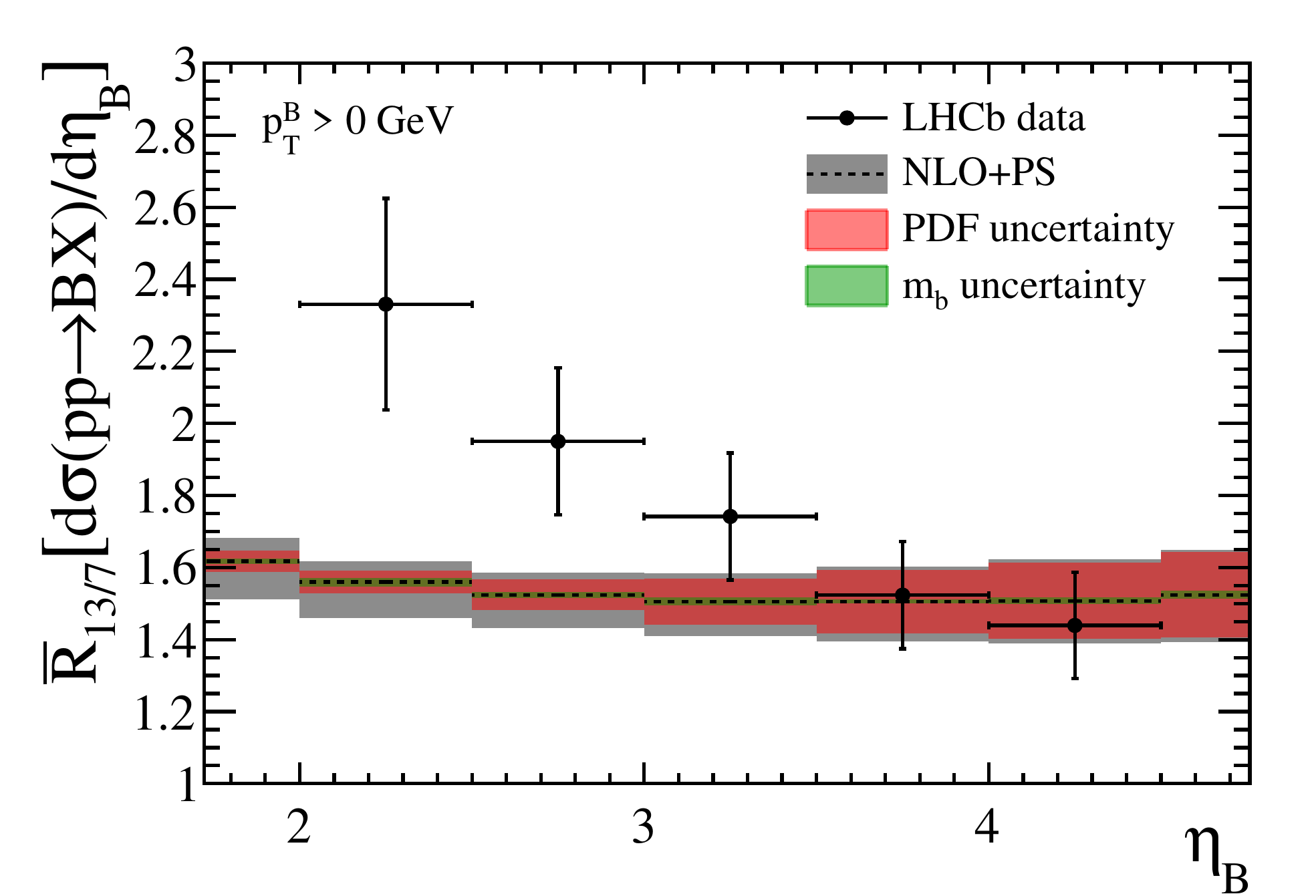}}
\makebox{\includegraphics[width=0.49\columnwidth]{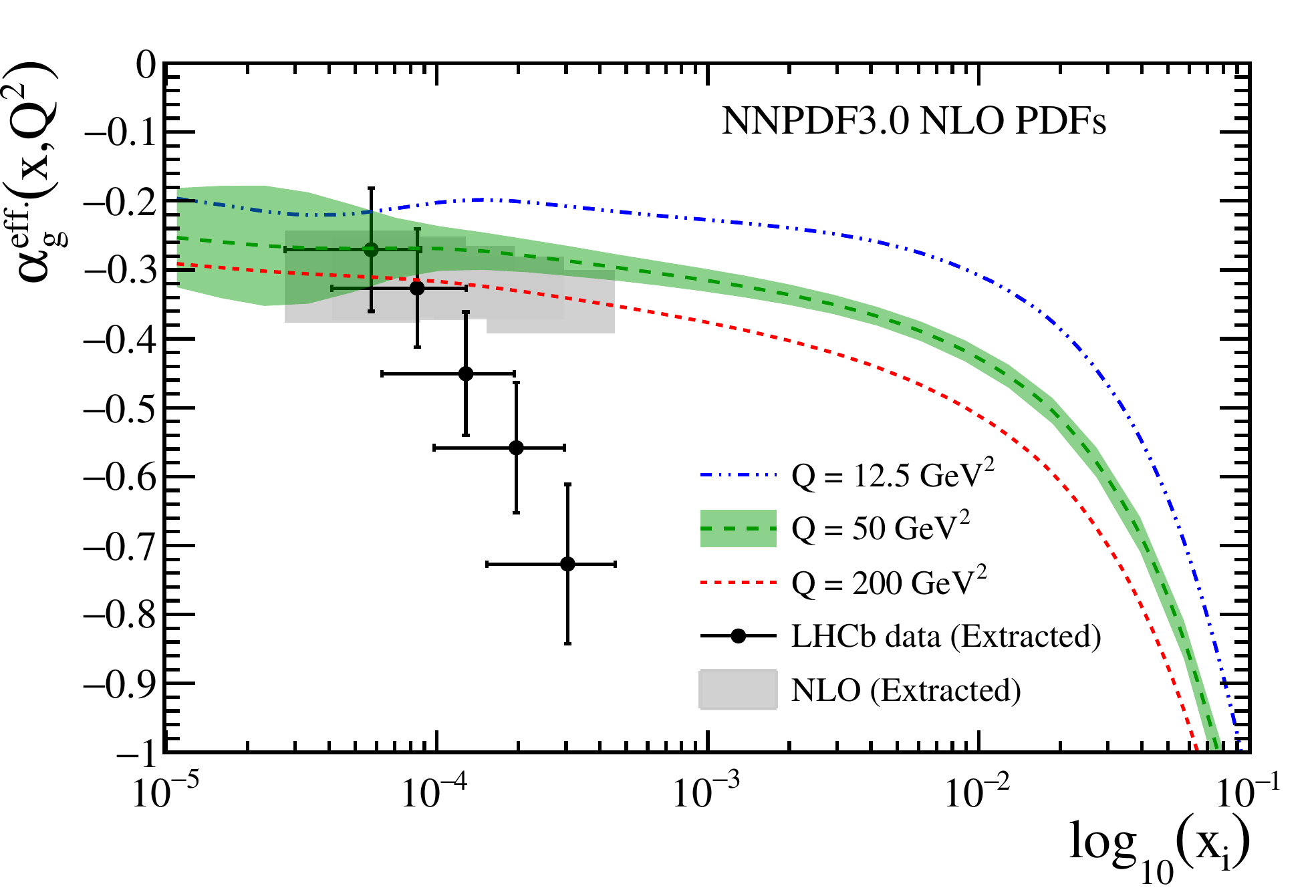}}
\end{center}
\vspace{0mm}
\caption{Left: The differential shifted ratio ($\overline{R}_{13/7}$) of inclusive B production with respect to pseudorapidity, 
measured within the LHCb acceptance. Right: The extracted values of $\alpha_{\rm g}^{\rm eff.}$ obtained from the LHCb 
data and NLO predictions, compared to those obtained directly from the input PDFs.}
\label{fig:RbarEta}
\end{figure}

A comparison of the LHCb data and the corresponding predictions of the pseudorapidity 
shifted ratio $\overline{R}_{13/7}$ are shown in Fig.~\ref{fig:RbarEta} (left). 
To obtain the experimental uncertainties, it is assumed that
the same strength of correlation between $\eta$-independent systematics 
quoted for the `non-shifted' ratio in Table~4 of~\cite{Aaij:2016avz} also applies to the shifted ratio.
In this case, the data is again observed to exceed the theoretical predictions
in the low pseudorapidity region of $\eta_B \in [2.0,3.0]$, and there is a clear
trend for the ratio to decrease with increasing $\eta_B$.

To understand this behaviour in terms of the low-$x$ gluon PDF, one can again consider 
the approximate relation between the shifted ratio and $\alpha_{\rm g}^{\rm eff.}$ introduced 
in Eq.~(\ref{eq:Extract}). In Fig.~\ref{fig:RbarEta} (right), the LHCb data has been extracted 
in a similar fashion to what was done for the rapidity shifted ratio in Fig.~\ref{fig:RbarApprox}. 
In this case, the $x$ dependence which enters the denominator of Eq.~(\ref{eq:Extract}) through 
$\bar{x}^{\rm 13~TeV}_2(\eta_{B}^{\prime})$ and $\bar{x}^{\rm 7~TeV}_2(\eta_{B})$ is extracted 
numerically in each pseudorapidity bin at LO. In addition, the flat `correction factor' of $1.05$ which 
accounts for the slight mis-alignment of the large-$x$ region is also applied. 
For reference, this method is also applied to the NLO prediction in exactly the same way.
The LHCb data clearly prefers large negative values of $\alpha_{\rm g}^{\rm eff.}$ around $x\sim3\cdot 10^{-4}$,
corresponding to an extremely fast growing gluon PDF, followed by a fast deceleration
in the growth of the gluon PDF at lower-$x$ values. The experimental and theoretical consistency
of this behaviour will be discussed in the following Section.

Before continuing, it is important to emphasise that the extraction of $\alpha_{\rm g}^{\rm eff.}$ in
this way is an approximation based on LO kinematics of the heavy quark production process.
Nevertheless, this approach is still extremely useful for studying the qualitative features of 
the data. In this case, demonstrating that a significant change in the behaviour of the low-$x$ gluon PDF
is necessary to accommodate the data.

\section{On the (in)consistency of the $B$ hadron data} \label{Consistency}
In the previous Section it was argued that, due to the kinematic alignment of the large-$x$ regions
present for the shifted ratio observable, the large deviation observed in data necessarily points to 
a significant modification of the behaviour of the low-$x$ gluon PDF.
The purpose of this Section is to discuss both the theoretical and experimental 
consistency of this behaviour.

\subsection{Theoretical consistency}
From inspection of Fig.~\ref{fig:RbarEta} (right), the LHCb data clearly prefers
large negative values of $\alpha_{\rm g}^{\rm eff.}$ in the region of $x\in [10^{-4},10^{-3}]$ and $Q^2 \sim {\rm50~GeV}^2$
which are inconsistent with the those obtained from global PDF fits.
This is a region where the shape of the gluon PDF is governed by a combination of both 
perturbative effects (through DGLAP evolution) and non-perturbative effects 
(through the input PDFs at $Q_0 \sim {\rm1~GeV}$).
To investigate the origin of the tension between the values of $\alpha_{\rm g}^{\rm eff.}$
extracted from the LHCb data and those obtained from global analyses of proton structure (see Fig.~\ref{fig:alpha}),
it is useful to introduce a toy model PDF set. Such an exercise is useful for understanding
the perturbative behaviour of $\alpha_{\rm g}^{\rm eff.}(x,Q^2)$ based on a simple
model for the structure of the non-perturbative inputs PDFs.

To do so, such a model is introduced into {\sc\small APFEL}~\cite{Bertone:2013vaa}
at the scale $Q_0 = {\rm1~GeV}$, based upon the following parametrisation of the input PDFs
\begin{align} \nonumber \label{eq:PDFs}
xq_V	(x)	&= N_{q_V} x^{\alpha_q} (1-x)^{\beta_q}	\,,\\ \nonumber
xS(x) 	&= N_{S} x^{\alpha_S} (1-x)^{\beta_S} 	\,,\\
xg(x) 	&= N_{g} x^{\alpha_{\rm g}} (1-x)^{\beta_g} \,.
\end{align}
The valence content $q_V(x)$ is practically implemented 
as $q_V(x) = 3/2 u_V(x) = 3 d_V(x)$. For the sea content $S(x)$
it is assumed that
\begin{align}
\bar{d}(x) = \bar{u}(x) = \bar{s}(x) = s(x) = (q_V(x)-q(x))/2 =  S(x)/6\,,
\end{align}
where $q(x) = u(x)+d(x)$. The form of the input PDFs is motivated
by non-perturbative QCD considerations~\cite{Regge:1959mz,Brodsky:1973kr}, 
where Regge theory predicts the low-$x$ behaviour $x^{\alpha}$, and Brodsky-Farrar 
quark counting rules predict the large-$x$ behaviour $(1-x)^{\beta}$. This functional form, now
superseded by much more flexible parameterisations, has long been the starting point
of PDF parameterisations. In the current model there are a total of 9 free parameters, 
two of which are fixed by the sum rules
\begin{align}
\int_0^1 dx\,q_V(x) = 3.0 \,, \qquad
\int_0^1 dx\,x\left( q_V(x)	+ S(x)+ g(x) \right) = 1.0 \,.
\end{align}
The first sum rule is used to fix the normalisation of the valence content ($N_q$), and the second 
the normalisation of the gluon PDF ($N_g$). The exponents of the valence quark content
are fixed to $\alpha_q = 0.5$ and $\beta_q = 3.0$, and it is found that altering these values 
has little impact on the qualitative behaviour of the low-$x$ gluon PDF.
As a benchmark, it is assumed the sea quark and gluon distributions
have identical shapes governed by $\alpha_{\rm g} = \alpha_S = -0.2$ and 
$\beta_g = \beta_S = 5.0$, with the normalisation $N_S = 3/4 N_g$.
Several variations of the benchmark model are then considered by 
enforcing a vanishing sea or gluon content at the starting scale $Q_0$.
Practically, these scenarios are achieved by setting either $N_S = 0$ or $N_g = 0$,
and correspond to generating the sea content or gluon PDF only perturbatively.
In addition, variations of the component $\alpha_{\rm g}$ 
are also considered, which modify both the shape of the gluon PDF at low-$x$
and also the normalisation (through the momentum sum rule).
The default choices for these parameters and the considered 
variations are provided in Table~\ref{tab:quark}.
\renewcommand*{\arraystretch}{1.0}
\begin{table}[h!]
	\centering
	\begin{tabular}{ c | c | c | c | c @{}}
      	\hline
	 			&	Default	&	$N_g = 0$ &	$N_s = 0$, $\alpha_{\rm g} = -0.4$ & $N_s = 0$, $\alpha_{\rm g} = -0.1$ \\ \hline
	 $\alpha_q$	&	0.5		&	0.5			&	0.5		&	0.5		\\
	 $\beta_q$	&	3.0		&	3.0			&	3.0		&	3.0		\\
	 $N_Q$		&	3.28		&	3.28			&	3.28		&	3.28		\\ \hline
	 $\alpha_S$	&	-0.2		&	-0.2			&	-		&	-		\\
	 $\beta_S$	&	5.0		&	5.0			&	-		&	-		\\
	 $N_S$		&	0.59		&	2.37			&	0		&	0		\\ \hline
	 $\alpha{\rm g}$	&	-0.2	&	-			&	-0.4		&	-0.1		\\
	 $\beta_g$	&	5.0		&	-			&	5.0		&	5.0		\\
	 $N_g$	 	&	1.77		&	0			&	1.29		&	3.11		\\ \hline
        \hline
	\end{tabular}
        \caption{Summary of the various input parameters for the considered toy model PDF sets.}
         \label{tab:quark}
\end{table}

\begin{figure}[ht!]
\begin{center}
\makebox{\includegraphics[width=0.49\columnwidth]{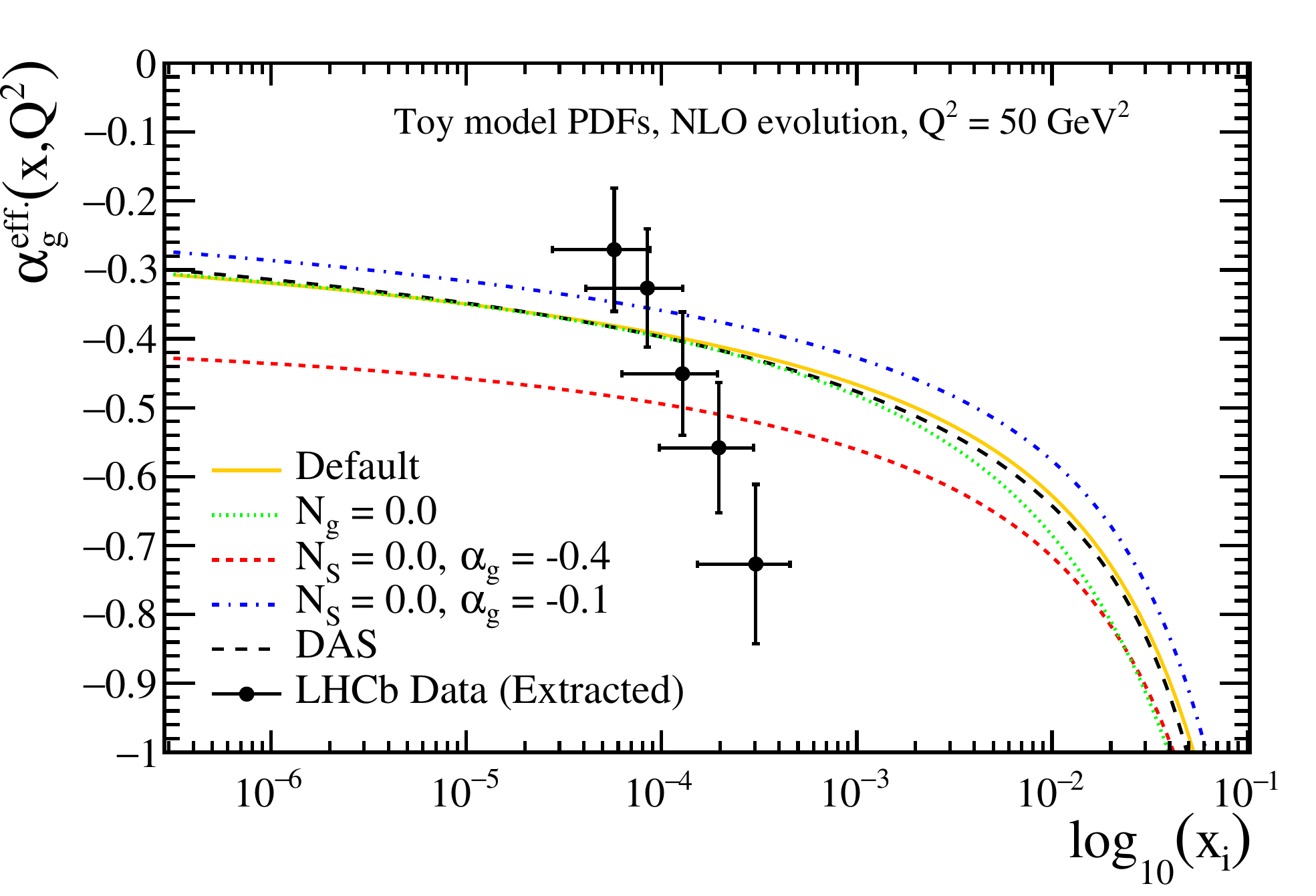}}
\end{center}
\caption{Predictions for $\alpha_{\rm g}^{\rm eff.}(x,Q^2)$ obtained from a toy PDF model defined
in Eq.~(\ref{eq:PDFs}) are compared to the extracted values from LHCb data.}
\label{fig:ToyPDFs}
\end{figure}

The perturbative behaviour of $\alpha_{\rm g}^{\rm eff.}(x,Q^2)$ is the examined 
in this model by evolving the PDFs at NLO QCD accuracy using the {\sc\small APFEL}
evolution routines. In a similar fashion to how $\alpha_{\rm g}^{\rm eff.}(x,Q^2)$
was computed for the LHAPDF grid files, the values of $xg(x,Q^2)$ obtained for the
toy models are tabulated on a grid in $x$ space and the derivate defined in Eq.~(\ref{eq:alpha}) is performed numerically.
The results of this study are shown in Fig.~\ref{fig:ToyPDFs} at the scale $Q^2 = {\rm 50~GeV}^2$, and
are compared to the extracted values from the LHCb data. In addition to the predictions
from the toy model variations, an analytic prediction based on the double asymptotic scaling of PDFs
(DAS)~\cite{Ball:1994du,Ball:1994kc,Forte:1995vs} is also shown. The logarithmic growth of the gluon 
PDF in this case is provided analytically, based on an approximation valid in the double limit of large-$Q^2$ and low-$x$, 
according to
\beq
\alpha_{\rm g}^{\rm DAS}(x,Q^2) = \frac{-1 + 4 \gamma \sigma}{4 \ln (x/x_0)} \,,  \quad
\sigma = \left[ \ln\left(\frac{x_0}{x}\right) \ln \frac{ \ln(Q^2/\Lambda^2)}{\ln(Q^2_0/\Lambda^2)}	\right]^{\frac{1}{2}} \,, \quad
\gamma = \left[	\frac{36}{33-6 n_f}\right]^{\frac{1}{2}}\,.
\eeq
The shown predictions are obtained with the input values $Q_0 = {\rm 1~GeV}$, $x_0 = 0.1$, $n_f = 5$ and
$\Lambda = 0.22{\rm~GeV}$\footnote{I thank Emanuele Nocera for a cross check of this implementation.}.

The same general features are found for $\alpha_{\rm g}^{\rm eff.}(x,Q^2)$ in all cases.
Firstly, the gluon PDF grows extremely quickly in the region $x\sim0.1$ as it is seeded by
is valence-like PDF content at large-$x$. This growth then decelerates with decreasing $x$ 
and eventually tends to a constant value, which depends on the choice input parameter 
$\alpha_{\rm g}$. The general behaviour of $\alpha_{\rm g}^{\rm eff.}(x,Q^2)$ for the toy model,
based on the simplified parameterisation in Eq.~\ref{eq:PDFs}, is therefore governed by a combination of the 
presence of valence-like PDFs at large-$x$ and DGLAP evolution effects. These are the same features which 
are also observed in global PDF fits --- see Fig.~\ref{fig:alpha}.

Based on these studies, it would seem the only way to accommodate the values of 
$\alpha_{\rm g}^{\rm eff.}(x,Q^2)$ preferred by the LHCb data, is to introduce
extremely ad-hoc behaviour in the non-perturbative gluon PDF in the range of $x\in[10^{-4}, 10^{-3}]$.
The reason is that the behaviour of $\alpha_{\rm g}^{\rm eff.}(x,Q^2)$ at large-$x$
is a general consequence of the valence-like content within the proton, which is well established. 
Therefore, to reach values of $\alpha_{\rm g}^{\rm eff.}(x\sim5\cdot10^{-4},Q^2 = {\rm 50~GeV}^2) \approx -0.7$
as indicated by the LHCb data, it is necessary to introduce a region of accelerated growth in the non-perturbative 
gluon PDF around $x \sim 10^{-3}$. This period of accelerated growth must then be closely followed by a period
of decelerated growth to accommodate the values of the ratio obtained at larger $\eta_B$ values.
Introducing such a feature into the definition of $g(x\sim10^{-3},Q_0)$ is in principle possible, since 
we should really be agnostic about the shape of non-perturbative object. However, the cost of doing
so would be to drastically change the predictions of many collider observables. As an example, 
consider the prediction of the inclusive charm and bottom Structure Functions $F_2^{qq}(x,Q^2)$, $q = c, b$,
which are an ingredient of the cross section prediction for charm and bottom production in DIS.
The LO prediction for this quantity is directly proportional to the gluon PDF, and is obtained
through the convolution of the gluon PDF with heavy quark coefficient function.
Measurements of both charm and bottom quark structure functions have been performed in
the range of $x \sim [10^{-4},10^{-3}]$ at $Q^{2}$ values of $6.5, 12.0, 25.0{\rm~GeV}^2$~\cite{Abramowicz:2014zub}.
No evidence for a steeply rising non-perturbative gluon PDF, which would result a sharp
rise of both $F_2^{cc}(x,Q^2)$ and $F_2^{bb}(x,Q^2)$, is observed.

\subsection{Experimental consistency with $D$ hadron data}
An another important consistency check can be performed by
drawing comparison to the available forward $D$ hadron data. The motivation for performing 
this check is that the theoretical framework for providing $B$ and $D$ hadron 
predictions is equivalent. In addition, there is LHCb data for $D$ hadron production 
in a kinematic regime which is highly correlated with that of $B$ hadron production\footnote{I am grateful to Michelangelo Mangano for 
this suggestion.}. Therefore, the consistency between $D$ hadron predictions and data provides an important 
cross check of the $B$ hadron results.

\begin{figure}[t!]
\begin{center}
\makebox{\includegraphics[width=0.49\columnwidth]{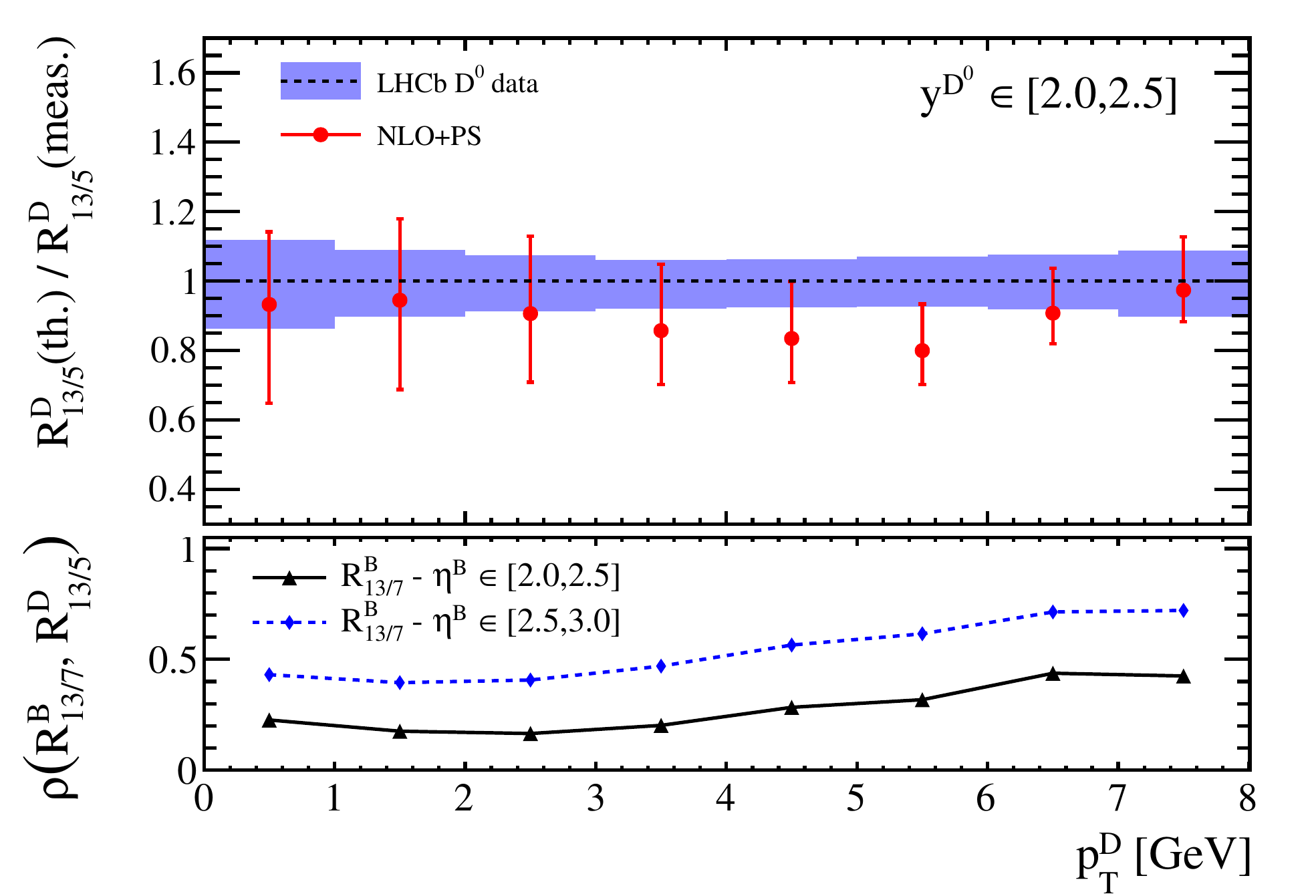}}
\makebox{\includegraphics[width=0.49\columnwidth]{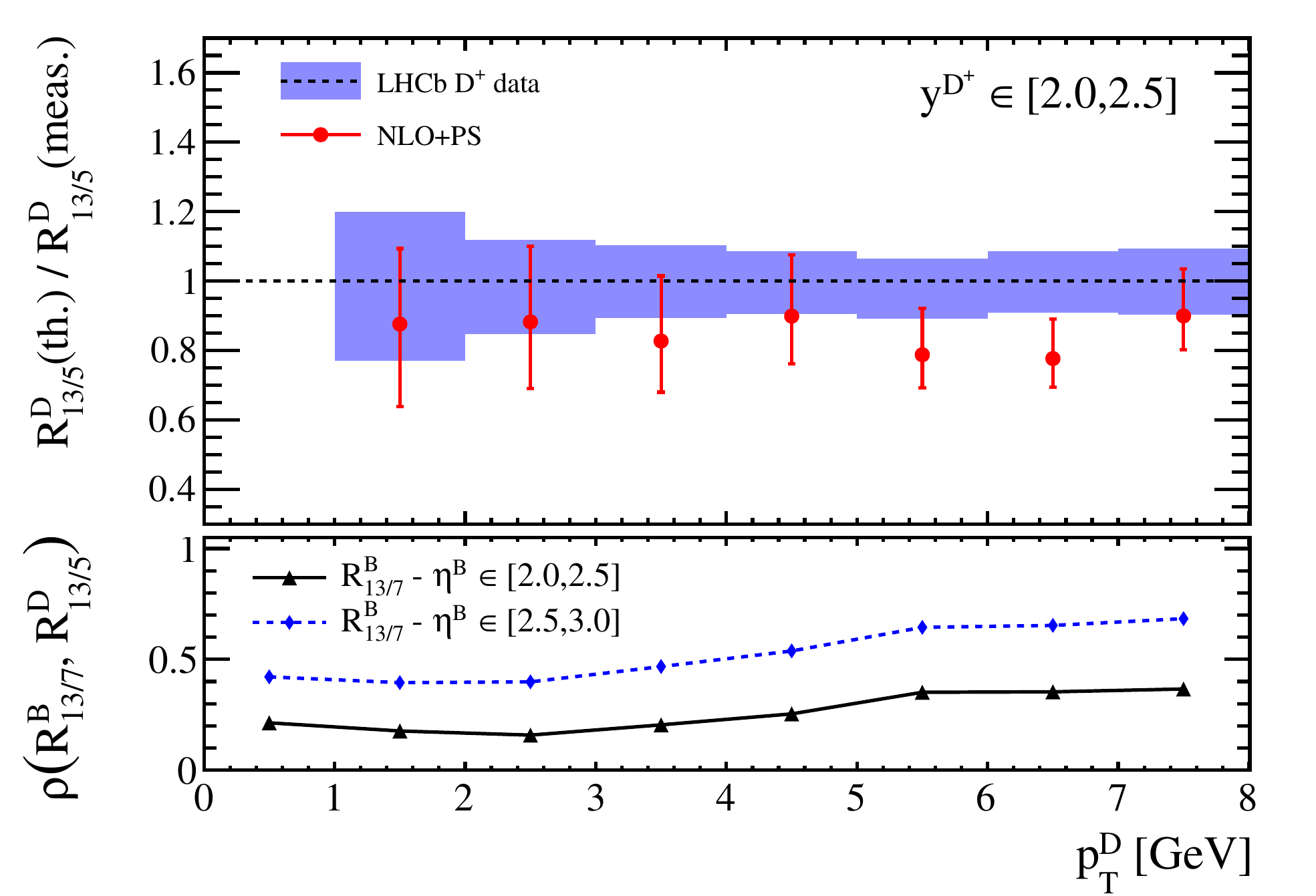}}
\end{center}
\caption{Comparison of the ratio of $D$ hadron cross section data at 13 and 5~TeV
within the kinematic region of $y_D \in [2.0,2.5]$, where both predictions and data
are normalised with respect to the central data point in each bin. In the lower
panel, the correlation of the $D$ and $B$ hadron ratio predictions are shown for specific kinematic selections.}
\label{fig:Corr135}
\end{figure}

\begin{figure}[ht!]
\begin{center}
\makebox{\includegraphics[width=0.49\columnwidth]{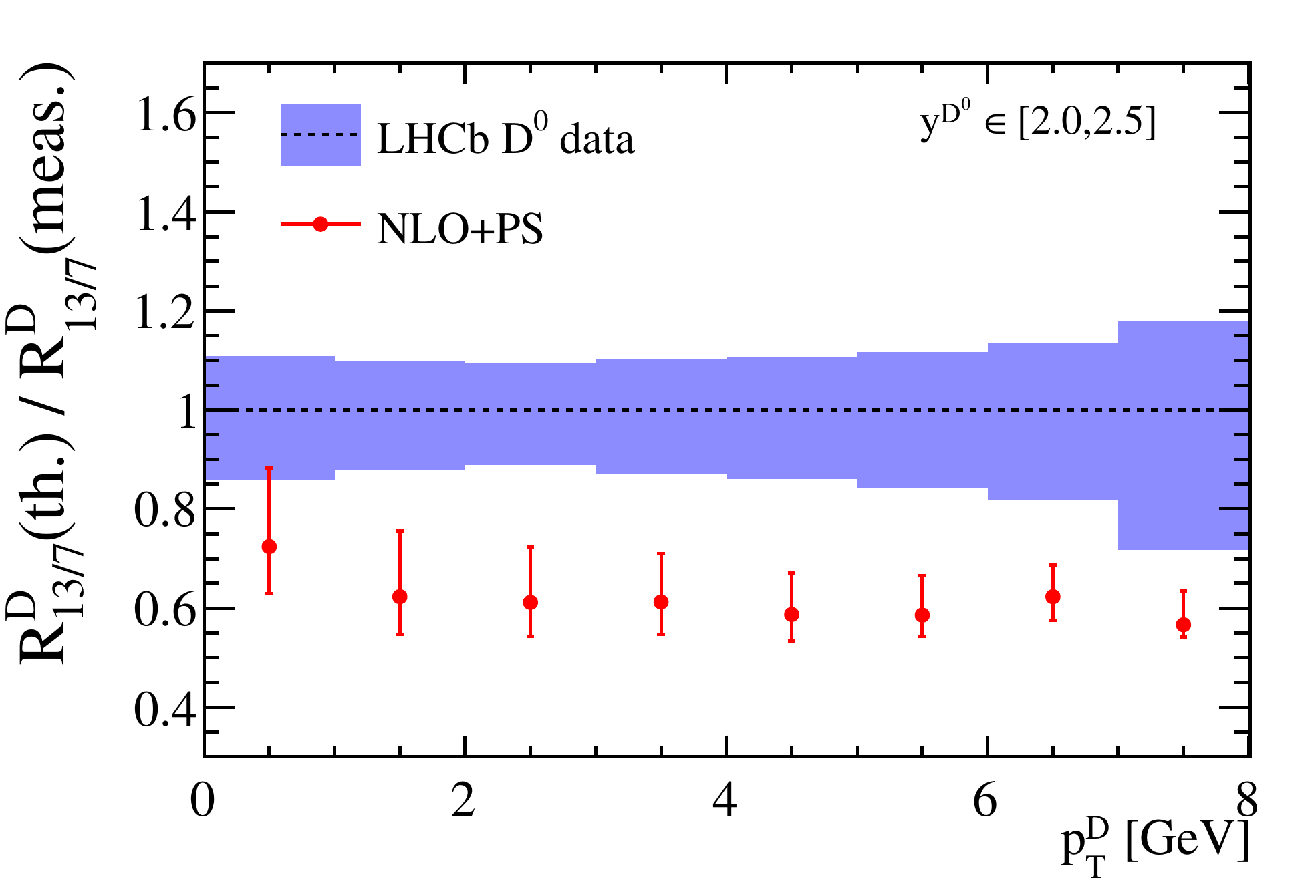}}
\makebox{\includegraphics[width=0.49\columnwidth]{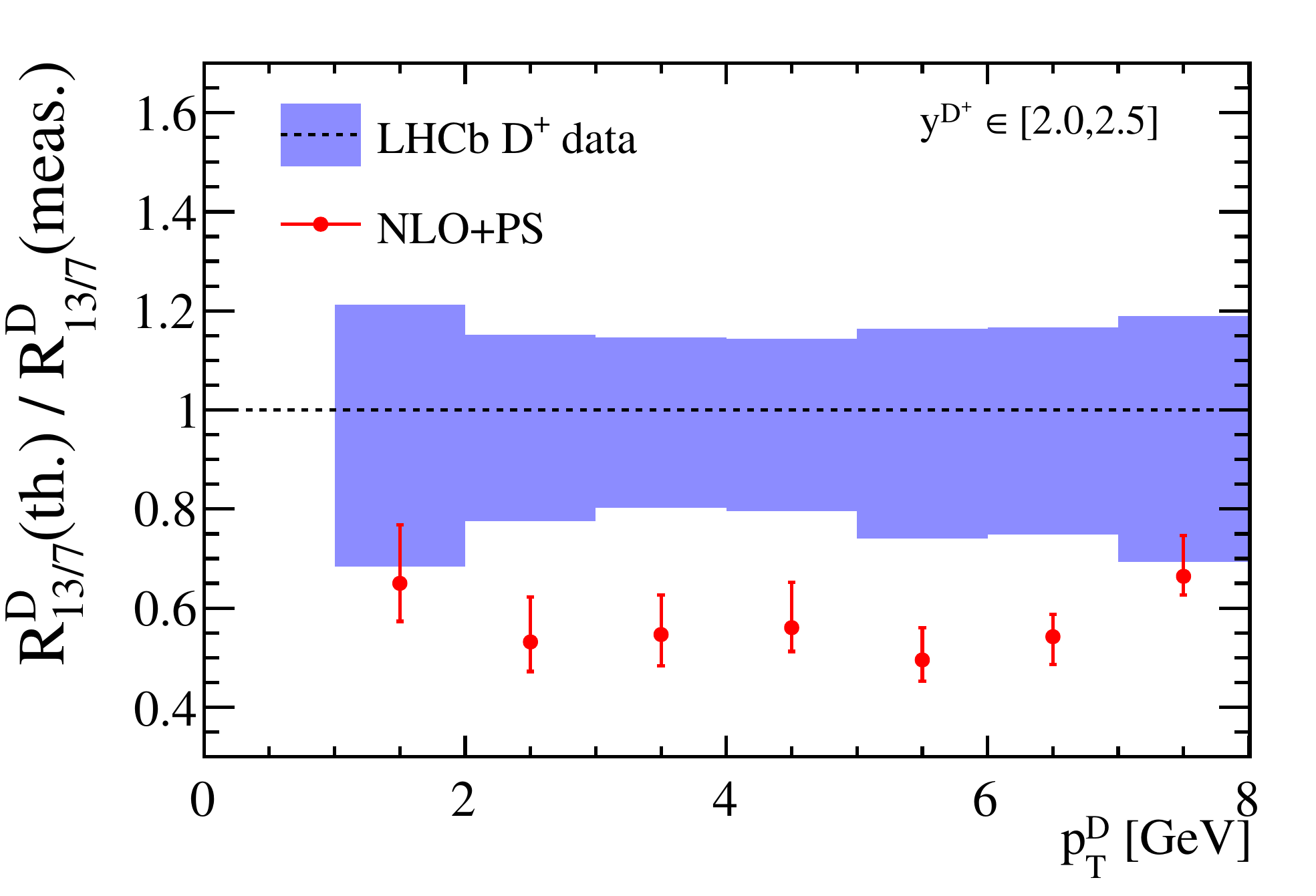}}
\end{center}
\caption{Same as Fig.~\ref{fig:Corr135} (upper), now for the ratio of $D$ hadron cross section data at 13 and 7~TeV.}
\label{fig:Corr137}
\end{figure}

Measurements of forward $D$ hadron production have been presented at 
5, 7, and 13~TeV~\cite{Aaij:2016jht,Aaij:2013mga,Aaij:2015bpa},
and as part of these measurements the ratio of double differential $D$ hadron production
at 13~TeV with respect to 5 and 7~TeV has been presented. 
These ratio measurements are available within the kinematic range of $y_D \in [2.0,4.5]$ and 
$p_T^D \in [0,8]$~GeV. In the following, comparisons of the $D$ hadron data are performed at the level
of the double differential ratio according to
\begin{align}
R_{Y/X}^D = \frac{d^2 \sigma^{\rm Y~TeV}(pp\to DX)}{d y^D_i  d (p_T^D)_j} \bigg{/} \frac{d^2 \sigma^{\rm X~TeV}(pp\to DX)}{d y^D_i  d (p_T^D)_j} \,.
\end{align}
As the tension in the $B$ hadron data is observed in the lower pseudorapidity bins, 
the focus of the $D$ hadron studies is also towards the low rapidity region
of $y^D \in [2.0,2.5]$. In addition, particular attention should be paid to the region
of $p_T^D \in [6,8]$~GeV, where similar values of $m_T$ are probed with respect to the $B$ hadron predictions. 
The first comparison is provided in the upper panel of Fig.~\ref{fig:Corr135} where the data and theoretical predictions
for $R_{13/5}^D$ are provided, normalised to the central value of the data. In the lower panel, the correlation
of the $D$ hadron ratio predictions with those of the $B$ hadron ratio predictions (for specific choices of pseudorapidity bins)
are shown.
As expected, the correlation between these predictions is strongest in the high $p_T$ range,
amounting to 0.4 and 0.7 for the $B$ hadron ratio in the pseudorapidity bins of $\eta_B \in [2.0,2.5]$ and
$\eta_B \in [2.5,3.0]$ respectively.  As shown in Fig.~\ref{fig:Ratio} (left), these are the two pseudorapidity bins 
for which the tension in data is observed, however the predictions of $R_{13/5}^D$ within
this region are entirely consistent with the data.
The same comparison is also performed for the experimentally less precise $R_{13/7}^D$ data, and
is shown in Fig.~\ref{fig:Corr137}. In this case, the measured ratio systematically exceeds the theoretical predictions.
This feature is exactly the same as that observed in the ratio of $B$ hadron cross section measurements
at 13 and 7~TeV, although the deviation is less significant in this case. 

\begin{figure}[ht!]
\begin{center}
\makebox{\includegraphics[width=0.49\columnwidth]{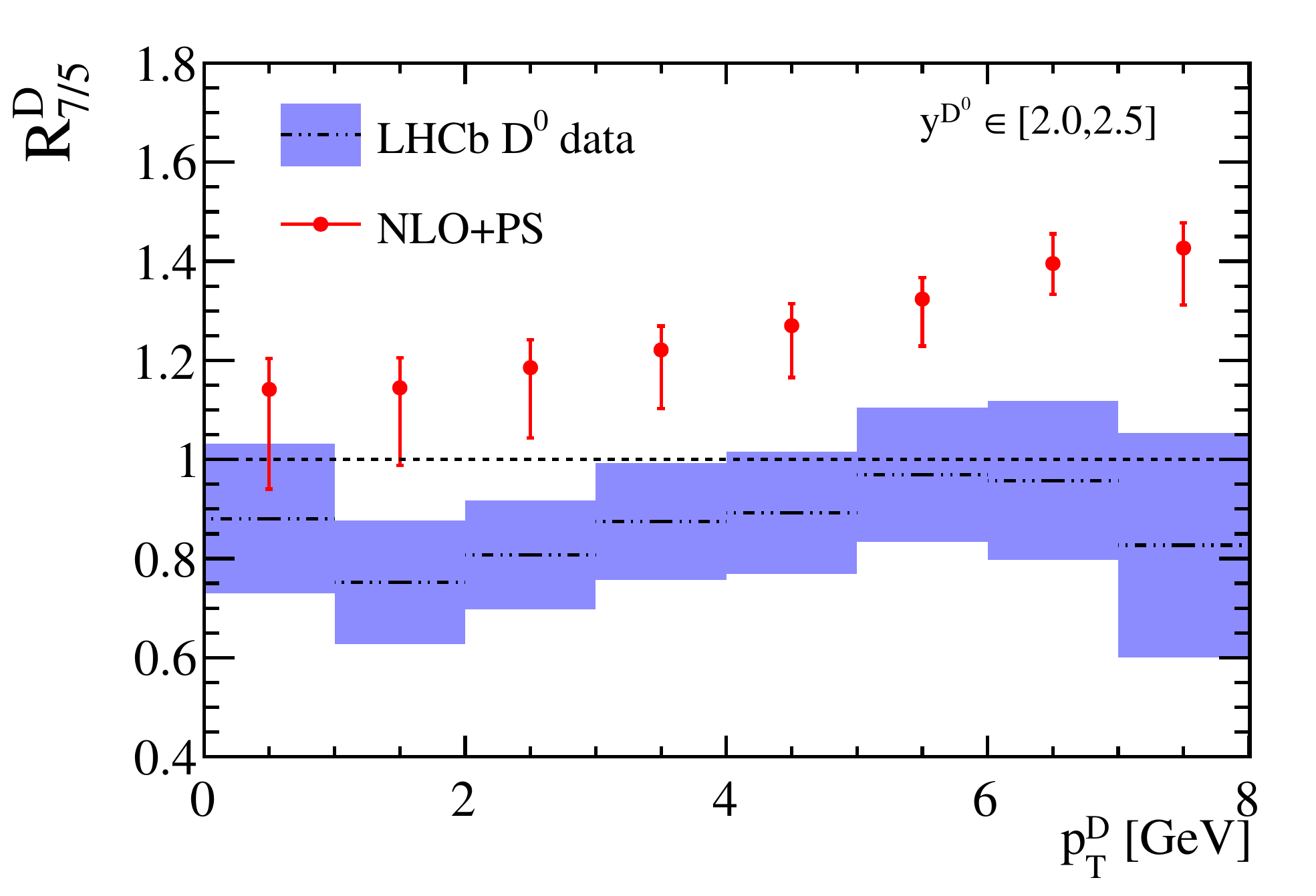}}
\makebox{\includegraphics[width=0.49\columnwidth]{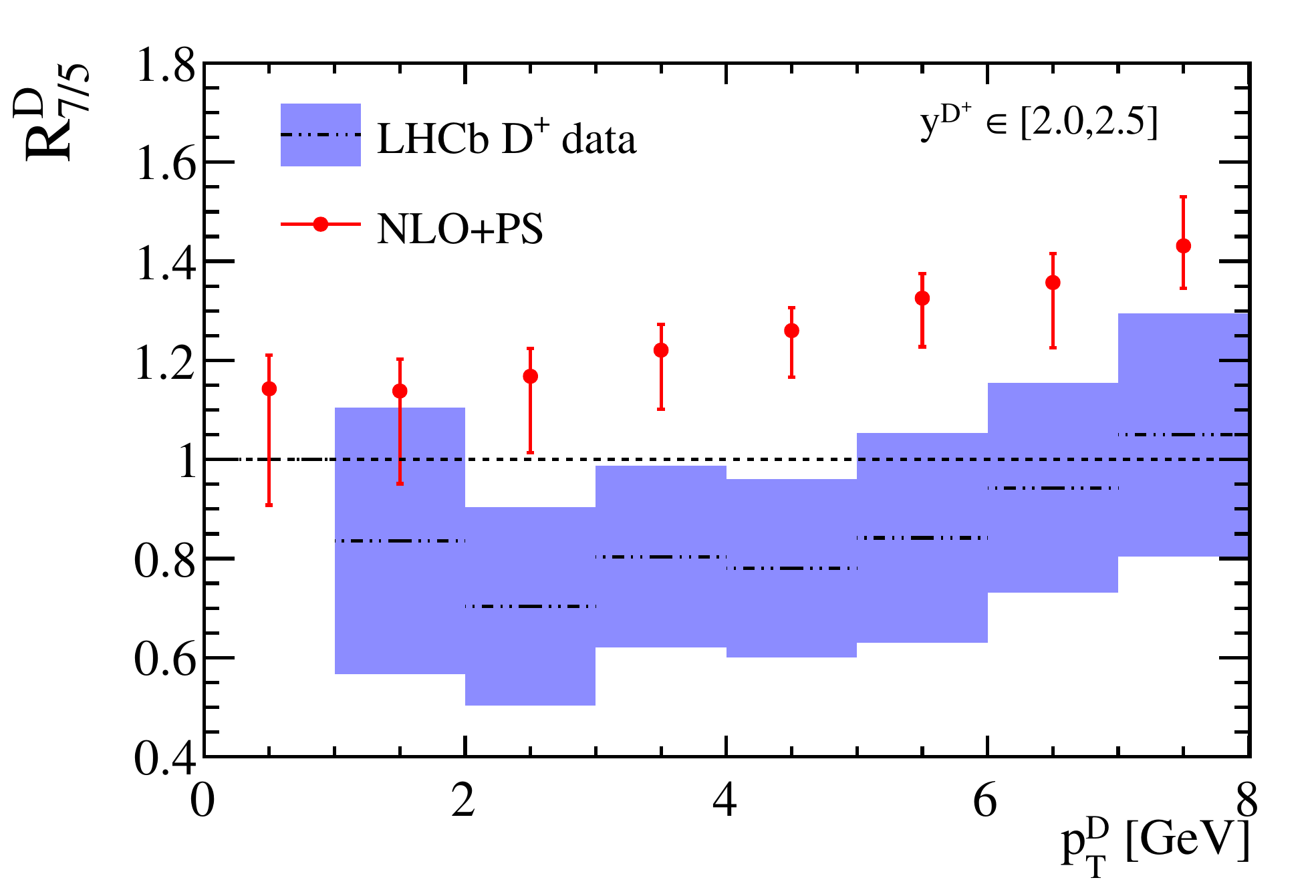}}
\end{center}
\caption{Comparison of the ratio of 7 and 5~TeV cross section data for $D$ hadrons
within the kinematic region of $y_D \in [2.0,2.5]$. Experimental uncertainties have been added in quadrature.}
\label{fig:Rat75}
\end{figure}

This situation is quite perplexing, as no deviation is found in the ratio of the 13 and 5~TeV cross section 
data, as shown in Fig.~\ref{fig:Corr135}, which is expected be more sensitive to changes in the shape of the gluon 
PDF both at small- and large-$x$ values. The fact that no deviation is observed in this case, suggests that the
$D$ hadron data is not self consistent.
Another way of viewing the tension in $D$ hadron data is to construct the ratio $R_{7/5}^D$ from
the available cross section data. This is done by adding the experimental uncertainties in quadrature 
(a direct measurement of this ratio was not presented), and the results of this combination are shown 
in Fig.~\ref{fig:Rat75} for the rapidity region of $y_D \in [2.0,2.5]$.
The experimental results for the ratio are generally below $1.0$, which
indicates that the differential cross section decreases with increasing CoM energy. These
results are not in line with expectations based on perturbative QCD, where the evolved gluon PDF
is expected to grow with decreasing $x$.

\section{Discussion and conclusions} \label{Conclusions}
The large discrepancy observed in the ratio of forward $B$ hadron production at 13 and 7~TeV
has motivated a detailed of study of the available LHCb data, both through normalised cross section
observables and various cross section ratios. It has been argued that, due to alignment of the PDF 
sampling regions, the tension present in the shifted pseudorapidity observable shown in 
Fig.~\ref{fig:RbarEta} (left) must be attributed to solely to the behaviour of the low-$x$ gluon PDF.
In fact, it is possible to approximately relate the deviation observed in data to the logarithmic
growth of the gluon PDF at low-$x$, described by the quantity $\alpha_{\rm g}^{\rm eff.}$ as defined
in Eq.~(\ref{eq:alpha}). After constructing this shifted ratio with the LHCb data, it is shown
that the extracted values of $\alpha_{\rm g}^{\rm eff.}$ are not consistent with the expectations
from global PDF fits --- see for example Fig.~\ref{fig:RbarEta} (right).
The reason for this tension is that the LHCb data indicates the presence of a region of accelerated 
growth of the gluon PDF, closely followed by a period of deceleration, within the kinematic
range of $x\in [10^{-3},10^{-4}]$. The only way to theoretically accommodate such behaviour is
to introduce this structure into the non-perturbative gluon PDF, since this sort of feature
is not generated by DGLAP evolution.
However, introducing this behaviour would also lead to an extremely fast growth of the
heavy quark structure functions $F_2^{QQ}(x,Q^2)$ within this $x$-range, which
is ruled out by measurements at HERA~\cite{Abramowicz:2014zub}.

Studies of the normalised cross section data at 7 and 13~TeV, as shown in Fig.~\ref{fig:dsigdeta7} and~\ref{fig:dsigdeta13} respectively, 
do not conclusively indicate a problem with a particular data set. The overall agreement with the data sets is reasonable, as quantified in Eq.~(\ref{eq:chi2}), while it is noted that there is local tension in the lowest pseudorapidity bin in the 7~TeV measurement of 2.1$\sigma$.
No such tension is observed for the 13~TeV measurement.
A further consistency check of the LHCb $B$ hadron data is performed by studying the ratios of forward $D$ hadron
production available at 5, 7 and 13~TeV. This study was focussed on the rapidity region of $y_D \in [2.0,2.5]$,
corresponding to the region where tension is observed for the $B$ hadron data. The measurement of
of the $13/5$~TeV $D$ hadron cross section ratio (which is experimentally most precise), is found to be fully consistent 
with the theoretical predictions. Of the available $D$ hadron ratio data, this observable is also the most correlated with the $B$ hadron
ratio, and is expected to be most sensitive to the shape of the low-$x$ gluon PDF.
However, a comparison to both $D$ hadron ratios of $13/7$~TeV and $7/5$~TeV indicate a similar tension
to what is observed for $B$ hadron ratio. That is, this ratio exceeds the theoretical expectations in the lower
(pseudo)rapidity region of $[2.0,2.5]$. In particular, the reported cross section for the 5~TeV cross section
measurement in this rapidity region is larger than that measured at 7~TeV. 

To summarise, systematic tension is observed in the LHCb cross section measurements of $B$ and $D$ hadrons
in the (pseudo)rapidity region of $[2.0,2.5]$. Based on consistency checks of the data (through ratios
and normalised distributions), this appears to be caused by a (pseudo)rapidity dependent efficiency
correction which affects either 7 or both 5 and 13~TeV cross section measurements.
If indeed this is the case, then the results analyses quantifying the impact of the LHCb $B/D$ hadron
data on proton structure~\cite{Zenaiev:2015rfa,Gauld:2015yia,Gauld:2016kpd} may also be affected.
It is worth pointing out that the PDF constraints from this data are strongest in the large rapidity region,
which seems to be a region which is least affected. Therefore, it is not likely that the results of these
analyses would qualitatively change.

An extraction of the low-$x$ gluon PDF obtained from analyses of forward $B$ and $D$ hadron 
requires reliable data. Given that a detailed understanding of both the magnitude and shape
of the gluon PDF below $x\sim10^{-5}$ has important consequences for a range of physics processes 
such as LHC (and future collider) phenomenology~\cite{Skands:2014pea,AbelleiraFernandez:2012cc,Mangano:2016jyj}, 
the predictions of atmospheric charm production~\cite{Gauld:2015kvh,Garzelli:2016xmx}, 
and the Ultra High Energy neutrino-nucleon cross section~\cite{Gauld:2016kpd,CooperSarkar:2011pa}. 
It is therefore vital that LHCb re-investigate the measurements of forward $B$ and $D$ hadron production.

\acknowledgments
I am grateful to Juan Rojo for previous collaboration and discussions relevant to this work.
In addition to Uli Haisch for motivating discussions at the start of this project.
I am also extremely grateful to my colleagues at ETH Z\"urich for many useful discussions 
relevant to this work.
The work of R.~G. is supported by the ERC Advanced Grant MC@NNLO (340983).

\bibliographystyle{JHEP}
\bibliography{HVQ}

\end{document}